\pdfoutput=1
\documentclass[twocolumn,english,aps,superscriptaddress,prb]{revtex4-1}

\usepackage{babel}
\usepackage{amsmath}
\usepackage{amssymb}
\usepackage{graphicx}
\usepackage{array}

\begin{document}

\title{Spontaneous dimerization, critical lines, and short-range correlations in a frustrated spin-1 chain}

\author{Natalia Chepiga}
\affiliation{Institute of Physics, Ecole Polytechnique F\'ed\'erale de Lausanne (EPFL), CH-1015 Lausanne, Switzerland}
\author{Ian Affleck}
\affiliation{Department of Physics and Astronomy, University of British Columbia, Vancouver, BC, Canada V6T 1Z1}
\author{Fr\'ed\'eric Mila}
\affiliation{Institute of Physics, Ecole Polytechnique F\'ed\'erale de Lausanne (EPFL), CH-1015 Lausanne, Switzerland}

\date{\today}
\begin{abstract} 
We report on a detailed investigation of the spin-1 $J_1-J_2-J_3$ Heisenberg model, a frustrated model with nearest-neighbor coupling $J_1$, next-nearest neighbor coupling
$J_2$, and a three site interaction $J_3\left[({\bf S}_{i-1}\cdot {\bf S}_i)({\bf S}_i\cdot {\bf S}_{i+1})+{\mathrm H.c.}\right]$ previously studied in [Phys. Rev. B 93, 241108(R) (2016)]. Using DMRG and exact diagonalizations, we show that the phase boundaries between the Haldane phase, the next-nearest neighbor Haldane phase and the dimerized phase can be very accurately determined by combining the information deduced from the dimerization, the ground-state energy, the entanglement spectrum and the Berry phase. By a careful investigation of the finite-size spectrum, we also show that the transition between the next-nearest
neighbor Haldane phase and the dimerized phase is in the Ising universality class all along the critical line. Furthermore, we justify the conformal embedding of the $SU(2)_2$ Wess-Zumino-Witten conformal field theory in terms of a boson and an Ising field, and we explicitly derive a number of consequences of this embedding for the spectrum along the $SU(2)_2$ transition line between the Haldane phase and the dimerized phase.
We also show that the solitons along the first-order transition line between the Haldane phase and the dimerized phase carry a spin-1/2, while those between
different dimerization domains inside the dimerized phase carry a spin 1. Finally, we show that short-range correlations change character in the Haldane and dimerized phases through disorder and Lifshitz lines, as well as through the development of short-range dimer correlations in the Haldane phase, leading to a remarkably rich phase diagram.
\end{abstract}

\maketitle


\section{Introduction}

\subsection{Background}
Antiferromagnetic Heisenberg spin chains have been studied intensively over the years. Adding frustration through competing interactions leads to a variety of new phases and quantum phase transitions. The most famous example is probably the $J_1-J_2$ spin-1/2 chain\cite{MajumdarGhosh} which undergoes spontaneous dimerization when the 
ratio of the next-nearest neighbor interaction to the nearest-neighbor one $J_2/J_1\simeq 0.2411$\cite{okamoto}. For the spin-1 chain, spontaneous dimerization has long
been known to be induced by a negative biquadratic interaction $J_\text{biq}$ exactly opposite to the bilinear one: $J_\text{biq}/J_1=-1$. The critical point is integrable with Bethe ansatz\cite{takhtajan,babujian}, and it is in the SU(2)$_2$ Wess-Zumino-Witten (WZW) universality class\cite{affleck86_1,affleck86_2,affleck_haldane} . Unlike in spin-1/2 chains however, a next-nearest neighbor interaction $J_2$ does {\it not} lead to dimerization, but induces a phase transition into a phase that consists of 
intertwined next nearest-neighbor (NNN) Haldane chains\cite{kolezhuk_prl,kolezhuk_connectivity}. More recently, it has been shown that a three-site interaction $J_3\left[({\bf S}_{i-1}\cdot {\bf S}_i)({\bf S}_i\cdot {\bf S}_{i+1})+{\mathrm H.c.}\right]$ that reduces to the next-nearest neighbor interaction for spin-1/2 is also able to induce a spontaneous dimerization in spin-S chains for arbitrary S, and that, at least up to $S=2$, the transition is in the SU(2)$_{2S}$ WZW universality class\cite{michaud1,michaud2}.

The combined effect of $J_2$ and $J_\text{biq}$ for the spin-1 chain has recently been investigated by Pixley et al\cite{nevidomskyy}, who came to the conclusion that the phase diagram only consists of the phases previously identified in the models with only one of these frustrating interactions ($J_2$ or $J_\text{biq}$): the Haldane phase, a spontaneously dimerized phase, and the NNN-Haldane phase. They also carefully investigated the short-range correlations, which become
incommensurate through Lifshitz and disorder transition line. The dimerization transition was argued to be either in the SU(2)$_2$ WZW universality class, or to be first order.

Shortly after, the combined effect of $J_2$ and of the three-site 
interaction $J_3$ has been studied by the present authors\cite{J1J2J3_letter}. The model is defined by the $J_1-J_2-J_3$ 
Hamiltonian:
\begin{multline}
\label{eq:H}
H=\sum_{i}\left(J_1{\bf S}_i\cdot {\bf S}_{i+1}+J_2{\bf S}_{i-1} \cdot {\bf S}_{i+1}\right)\\
+\sum_{i}J_3\left[({\bf S}_{i-1}\cdot {\bf S}_i)({\bf S}_i\cdot {\bf S}_{i+1})+{\mathrm H.c.}\right]
\end{multline}
The phases turn out to be the same as for the $J_1-J_2-J_\text{biq}$ model, but, quite surprisingly, the dimerization transition between the NNN-Haldane and the dimerized phase was found to be in the Ising universality class. 
The goal of the present paper is to give a detailed account of how these conclusions were reached for the $J_1-J_2-J_3$ model, and to investigate the nature of the short-range correlations, which were not touched upon in the previous paper. 
The apparent discrepancy regarding
the nature of the transition from the NNN-Haldane to the dimerized phase between the $J_1-J_2-J_\text{biq}$ and the $J_1-J_2-J_3$ model has been resolved
since then by the present authors\cite{J1J2Jb_comment}, and the transition appears to be in the Ising universality class in both cases.

\subsection{Previous results}

The spin-1 chain with isotropic nearest-neighbor Heisenberg ($J_2=J_3=0$) interaction has been shown to be gapped \cite{Haldane} with exponentially decaying spin-spin correlations. The system is topologically non trivial, and the ground state has a hidden order characterized by a non-local string order parameter. For open boundary conditions, spin-1/2 edge states appear and give rise to two low-lying states, a singlet and the so-called Kennedy triplet\cite{Kennedy}. More recently, it was shown that the Haldane phase is characterized by the double degeneracy of the entanglement spectrum \cite{pollmann}. This degeneracy is protected by the same set of symmetries that protect the stability of the Haldane phase. If  the Hamiltonian is deformed while preserving these symmetries, the degeneracy can be lifted only by crossing a phase boundary.

The model with $J_3=0$ has been studied using a variational ansatz and the density matrix renormalization group (DMRG)\cite{kolezhuk_prl,kolezhuk_prb,dmrg1,dmrg2}. The authors have shown that the Haldane phase is stable until $J_2=0.7444(6)$, where a phase transition to the NNN-Haldane phase takes place. According to DMRG calculations, the spin gap remains open. The finite jump in the string order parameter suggests that the phase transition is first order, although no discontinuity could be identified in the first derivative of the energy at the transition. Disorder and Lifshitz points (i.e. points, where the correlation function in real space becomes incommensurate with a wave-vector $q\neq 0,\pi/2,\pi$, or where the structure factor has two peacks at $q\neq 0,\pi/2,\pi$, respectively) were identified at $\alpha_d=0.284(1)$ and $\alpha_L=0.3725(25)$. 

For the model with $J_2=0$, there is a transition at $J_3\simeq0.111$\cite{michaud1} to a spontaneously dimerized phase. This transition is continuous and belongs to the $\mathrm{SU}(2)_{k=2}$ WZW universality class\cite{affleck_haldane}.

There is also a line where the ground state is known exactly. Michaud et al.\cite{michaud1,michaud2} have shown that there is an exactly dimerized point for all spin-S chains for the $J_1-J_3$ model at $J_3/J_1=1/(4S(S+1)-2)$. For spin-1/2, this model reduces to the $J_1-J_2$ model with $J_2=J_3/2$, and this exactly
dimerized state can be seen as the generalization of the Majumdar-Ghosh point of the spin-1/2 $J_1-J_2$ chain. Further investigations have shown that
this result can be extended to the case where a next-nearest neighbor exchange $J_2$ interaction is included\cite{wang}. Indeed, provided that
\begin{equation}
\frac{J_3}{J_1-2J_2}=\frac{1}{4S(S+1)-2}.
\label{nnn_dimerization}
\end{equation}
the two fully dimerized states are eigenstates, and they are ground states if $J_2$ is not too large. 
Now, for $J_3=0$ and $S=1$, it has already been shown by Roth and Schollw\"ock that the ground state is not dimerized for $J_2=1/2$, but that it
lies in the Haldane phase\cite{kolezhuk_prl,kolezhuk_prb}.
 This suggests that, for spin 1, the transition between the dimerized phase and the Haldane phase, which is continuous
for $J_2=0$, has to become first order somewhere on the line $J_2+3J_3=1/2$ 
(the form taken by the condition of Eq.~\ref{nnn_dimerization} for $S=1$).

The phase diagram of the $J_1-J_2-J_3$ model (Fig.\ref{fig:PD}) was reported previously in Ref.\onlinecite{J1J2J3_letter}, in which we have discussed in details the nature of the phase transitions into the spontaneously dimerized phase. In particular, it was shown that the phase transition between the Haldane and dimerized phases is either WZW $\mathrm{SU}(2)_{k=2}$ or first order depending on the value of the coupling constant of the marginal operator, while the transition between the NNN-Haldane and dimerized phases is in the Ising universality class. In addition, we have suggested that the type of continuous transition depends on the nature of the domain walls between the phases: the transition is magnetic (WZW $\mathrm{SU}(2)_{k=2}$) if the domain wall carries a free spin, while it occurs in singlet sector and is in the Ising universality class otherwise.

\subsection{Scope}

In this paper, we report on an in-depth numerical and analytical investigation of the model of Eq.\ref{eq:H} using DMRG, exact diagonalizations, and field theory. Without loss of generality, we set $J_1=1$ throughout the paper, and we concentrate on the antiferromagnetic case $J_2\geq0$ and on positive three-site interaction $J_3\geq0$. In particular, we discuss the dimerization, the groundstate energy and the entanglement spectrum, all obtained by a matrix product state implementation of DMRG known as variational MPS\cite{dmrg3,dmrg4}, and the Berry phase\cite{berry,hatsugai} calculated with exact diagonalization.
We also confirm the magnetic nature of the domain walls between the Haldane and dimerized phases by looking at the solitons at the first order transition between these phases.
Furthermore, coming back to the nature of the phase transition between the NNN-Haldane and dimerized phases, we provide numerical evidence that the universality class is Ising all along the critical line, including the triple point. Finally, we discuss a variety of short range phases that appear on top the main phases.

The paper is organized as follows. We start with a brief discussion of the phase diagram in section \ref{sec:phase_diagram}. Section \ref{sec:CFTandDMRG} discusses the  conformal embedding used in the field theory approach and provides some technical  details on DMRG calculations. In section \ref{sec:methods}, we describe in more details how the phase diagram was obtained by a careful investigation of the dimerization order parameter, of the energy, of the entanglement spectrum, and of two types of Berry phases.  In section \ref{sec:solitons} we discuss solitons that appear at the first order phase transition between the Haldane and dimerized phases. Section \ref{sec:ising} discusses the limits of the Ising critical line: the triple point and the $J_2-J_3$ model. Section \ref{sec:short_range_order} gives additional details about the short-range orders realized in the system. We conclude with a summary of our main results in section \ref{sec:conclusion}.


\section{Summary of main results}
\label{sec:phase_diagram}

\subsection{Phases and transitions}

\begin{figure}[h!]
\includegraphics[width=0.47\textwidth]{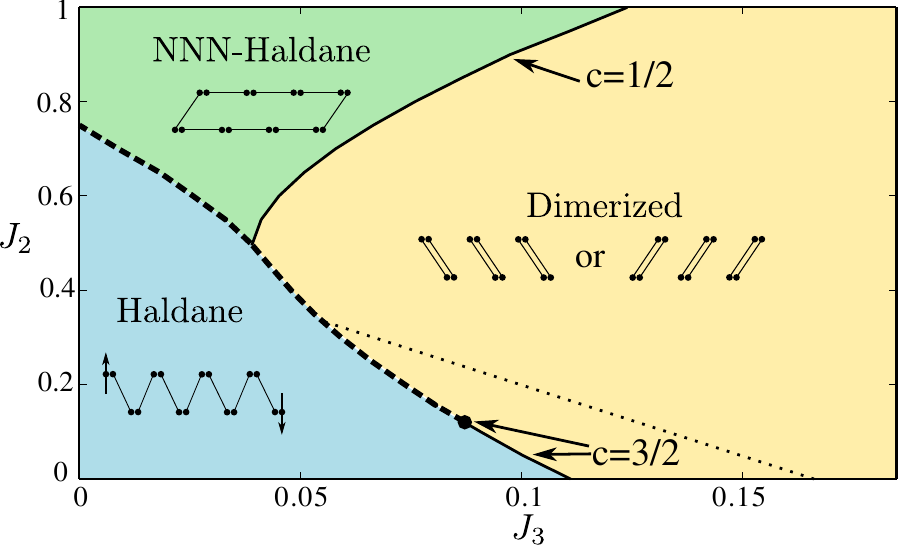}
\caption{(Color online) Phase diagram of the one-dimensional, spin-1 $J_1 - J_2 - J_3$ model. The transition from the dimerized phase to the Haldane phase is continuous along the solid line and first order along the dashed line. The transition from the NNN-Haldane to the dimerized is always continuous. The transition between the Haldane phase and the NNN-Haldane phase is always first order. The dotted line is the line where the ground state is exactly dimerized}
\label{fig:PD}
\end{figure}

Our main results are summarized in the phase diagram of Fig.\ref{fig:PD}. It consists of three phases: a Haldane phase around the nearest-neighbor Heisenberg chain ($J_2=J_3=0$), a next-nearest neighbor (NNN)-Haldane phase upon increasing $J_2$, and a dimerized phase upon increasing $J_3$. 
Each phase has a simple valence-bond-solid (VBS) representation sketched on the diagram. In this representation, each spin-1 is represented as a pair
of spins 1/2, and bonds correspond to spin singlets built out of two spins 1/2.

The transition between the Haldane phase and the dimerized phase starts at $J_2=0,J_3\simeq 0.111$, and it remains continuous along a line
up to the point  $J_2\simeq0.12,J_3\simeq 0.087$. On this line, and at the end point, the transition is characterized by a central charge $c=3/2$ and is in the $SU(2)_{k=2}$ WZW universality class. Beyond, the transition is first order.

The transition between the Haldane phase and the NNN-Haldane phase is first order. It is a topological transition: the two phases cannot be distinguished
by any local symmetry, but Haldane phase is topological with gapless edge excitations (so-called Kennedy triplet \cite{Kennedy}), whereas the NNN-Haldane phase is not. 
One can also resort to the non-local string order parameter, or to probes of topological properties such as the entanglement spectrum or the Berry phase, to distinguish them. 

Finally, the transition between the NNN-Haldane phase and the dimerized phase is in the Ising universality class with a central charge $c=1/2$. The singlet-triplet gap does not close at this transition.

On top of these transition lines, there is a remarkable line $J_2+3J_3=1/2$ along which the fully dimerized state is an exact eigenstate of the model.
This state is the ground state until the point $J_2\simeq0.335,J_3\simeq0.055$. At that point, the system undergoes a strongly first order
transition into the Haldane phase.

\subsection{Short-range order}

In addition to these phases, which can be distinguished by their topological properties or by the development of long-range dimerization, we have also identified regions of the phase diagrams characterized by various types of short-range order. The discussion and notations follow closely those of 
Ref.\onlinecite{nevidomskyy}, in which a very detailed investigation of the same problem for the  $J_1-J_2-J_\text{biq}$ model has been reported. The 
lack of long-range order to distinguish these regeions prevents them from being true phases in the thermodynamic sense. However, they play an important role in understanding the evolution of correlations in the phase diagram, and we will nevertheless refer to them as phases.

The correlation function $C(x)=\langle{\bf S}(0)\cdot{\bf S}(x)\rangle$ can be well accounted for throughout by the product of the two-dimensional Ornstein-Zernicke (OZ) form:
\begin{equation}
C_{OZ}(x)\propto\cos(q\cdot x)\frac{e^{-x/\xi}}{\sqrt{x}},
\label{eq:OZ}
\end{equation}
with, in some cases, a prefactor $1+\delta(-1)^x$, leading to the dimerized Ornstein-Zernicke (DOZ) form:
\begin{equation}
C_\mathrm{DOZ}\propto (1+\delta(-1)^x)C_{OZ}(x),
\label{eq:OZD}
\end{equation}
The wave number $q$, the correlation length $\xi$, and the dimerization parameter $\delta$ are fitting parameters that depend on the couplings $J_2$ and $J_3$. 

Note that the same form applies to the dimerized and non-dimerized phases, except, of course, a line of continuous WZW SU$(2)_2$ phase transition, at which the spin-spin correlation decays algebraically $C(x)\propto (-1)^x/|x|^{3/4}$ up to logarithmic corrections. The dimerized phase is characterized by the
development of long-range correlations of the two-spin operator ${\bf S}_i\cdot {\bf S}_{i+1}$.

The structure factor is defined by the Fourier transform of real space correlations $\langle{\bf S}_i\cdot{\bf S}_j\rangle$:
\begin{equation}
SF(q)=\frac{1}{N}\sum_{i,j}e^{iq(i-j)}\langle0|{\bf S}_i\cdot{\bf S}_j|0\rangle
\label{eq:sf}
\end{equation}

\begin{figure}[h!]
\includegraphics[width=0.45\textwidth]{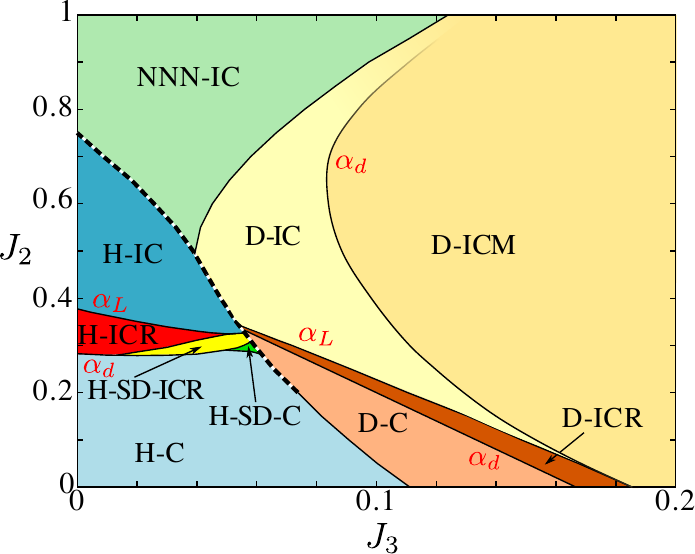}
\caption{(Color online) Phase diagram based on the type of short-range order realized in the phases of Fig.\ref{fig:PD}.  The notations for the different phases are described in the text.  Note that the disorder line $\alpha_d$ is distinct from the Lifshitz line $\alpha_L$ in both Haldane and dimerized phases. The line where the fully dimerized state is an exact ground-state coincides with the disorder line $\alpha_d$ in the dimerized phase.}
\label{fig:PD_short}
\end{figure}

Various short-range commensurate and incommensurate phases are shown in Fig.\ref{fig:PD_short}. Below we provide a short description of each phase. The detailed discussion of the form of the correlations that led to the identification of short-range order can be found in Section \ref{sec:short_range_order}.\\

{\bf Haldane Phase}
\begin{itemize}
\item{{\bf H-C:} Short-range antiferromagnetic order with commensurate real-space correlation function. $C(x)$ is well described by the OZ form with 
$q=\pi$ and no dimerization ($\delta=0$). The structure factor $SF(q)$ has a single peak at $q=\pi$.}
\item{{\bf H-SD-C:} Short-range dimer phase with commensurate real-space correlations ($q=\pi$). $C(x)$ is well described by the dimerized OZ form with 
$q=\pi$ and $\delta>0$. $SF(q)$ has a single peak at $q=\pi$.}
\item{{\bf H-SD-ICR:} Short-range dimer phase with incommensurate real-space correlations, characterized by $q>\pi$ and $\delta> 0$ in Eq.\ref{eq:OZD}. $SF(q)$ has a single peak at $q=\pi$.}
\item{{\bf H-ICR:} Short-range antiferromagnetic order with incommensurate real-space correlations. $C(x)$ is well described by the OZ form with 
$q\neq\pi$ and no dimerization ($\delta=0$). $SF(q)$ has a single peak at $q=\pi$. }
\item{{\bf H-IC:} Short-range antiferromagnetic order with incommensurate correlations in both real and momentum spaces. $C(x)$ is well described by the OZ form with $q>\pi$ and no dimerization ($\delta=0$), but $SF(q)$ has two symmetric peaks at $q\neq\pi$.}
\end{itemize}

{\bf Dimerized phase}
\begin{itemize}
\item{{\bf D-C:} The spin correlations are commensurate in both, real and momentum space. $C(x)$ is well described by the dimerized DOZ form with $q=\pi$ and $\delta>0$. $SF(q)$ has a single peak at $q=\pi$.}
\item{{\bf D-ICR:} Real-space correlations are incommensurate, and $C(x)$ is well fitted by the DOZ form with $\delta>0$ and $q>\pi$. $SF(q)$ still has a single peak at $q=\pi$.}
\item{{\bf D-IC:} The spin correlations are incommensurate in both real and momentum space, $C(x)$ is well fitted by the DOZ form with $\delta>0$ and $q>\pi$, but $SF(q)$ has two symmetric peaks at $q\neq\pi$.}
\item{{\bf D-ICM:} The spin correlations are incommensurate in momentum space, $SF(q)$ has two symmetric peaks at $q\neq\pi,\pi/2$. Real space correlations are commensurate with $q=\pi/2$}.
\end{itemize}

{\bf NNN-Haldane phase}
\begin{itemize}
\item{{\bf NNN-IC:}  The spin-spin correlations are incommensurate in both, real and momentum space. $C(x)$ is reasonably well fitted by the OZ form  with $q>\pi$ and no dimerization ($\delta=0$). $SF(q)$ has two symmetric peaks at $q\neq\pi$.}
\end{itemize}


\section{Methods}
\label{sec:CFTandDMRG}
\subsection{Conformal Embedding}
An important technique used in [\onlinecite{J1J2J3_letter}] was a conformal embedding - an exact equivalence of the $SU(2)_2$ WZW 
conformal field theory (CFT) with the direct product of Ising and free boson  CFT's. This was convenient since both sectors are gapless along the second 
order transition line between Haldane and dimerized phases while only the Ising sector is gapless along the transition line between NNN-Haldane and 
dimerised phases. The conformal embedding implies that each operator in the $SU(2)_2$ field theory can be written as a product of operators in the 
Ising and free boson theories. It also implies that the finite size spectra  are identical once certain selection rules are imposed. Here we give those selection rules, 
which were used to derive Table II of the Supplementary Material in  [\onlinecite{J1J2J3_letter}]. We consider the case of open boundary conditions (OBC) 
on the spin chain with $N\gg 1$ sites, at the tri-critical point. The $SU(2)_2$ WZW model has 3 conformal towers, labelled by lowest energy states of spin 
$j=0$, $1/2$ and $1$. The states in the spin-$j$ conformal tower have energies:
\begin{equation}
E_{j,n}={\pi v\over N}\left[-{1\over 16}+{j(j+1)\over 4}+n\right]
\end{equation}
for non-negative integer $n$. Excited states form multiplets of arbitrary spin, subject to the condition of being integer for $j=0$, $1$ and half-integer for $j=1/2$. In general,
multiplets of a given spin and energy occur with integer degeneracies $\geq 1$. As argued in [\onlinecite{J1J2J3_letter}],  the low energy 
spectrum of the spin chain at the tri-critical point is given by the $j=0$ conformal tower  for $N$ even and the $j=1$ conformal tower for $N$ odd. 
The Ising model has 3 conformal towers with energies
\begin{equation} E_\text{Ising}={\pi v\over N}\left[-{1\over 48}+x^{\text{Ising}}\right]
\end{equation}
where 
\begin{equation}x^{\text{Ising}}=x+n;\end{equation}
 $x=0$, $1/2$ or $1/16$, corresponding to the $I$, $\sigma$ and $\epsilon$ conformal towers and $n$ is a non-negative integer. The complete fss of the free boson model is
\begin{equation}
E_\text{boson}={\pi v\over N}\left[ -{1\over 24}+{(S^z)^2\over 2}+\sum_{n=1}^\infty m_nn\right]\label{Eb}
\end{equation}
where $S^z$ labels the quantum number of the state and the $m_n$ are non-negative integers.  This follows from the standard mode 
expansion for a periodic boson with $S^z$ the winding mode and the $m_n$ labelling excitations of the oscillator modes. 
Note that unlike the $SU(2)_2$ and Ising cases, we have a simple explicit formula for all energies in the free boson case, labelled by $S^z$. 

\begin{table}[h]
\centering 
\begin{tabular}{c|c|c|c|c}
n&$S^z$&$m_1$&$m_2$&$x^\text{Ising}$\\
\hline \hline 
0&0&0&0&0\\
\hline
1&$\pm 1$&0&0&1/2\\
1&0&1&0&0\\
\hline 
2&$\pm 2$&0&0&0\\
2&$\pm 1$&1&0&1/2\\
2&$\pm 1$&0&0&3/2\\
2&0&2&0&0\\
2&0&0&1&0\\
2&0&0&0&2
\end{tabular}
\caption{Details of $j=0$ conformal tower.}\end{table}

\begin{table}[h]
\centering
\begin{tabular}{c|c|c|c|}
n&s=0&s=1&s=2\\
\hline 
0&1&0&0\\
1&0&1&0\\
2&1&1&1
\end{tabular}
\caption{j=0 Conformal Tower, labelled by multiplicities of spin multiplets at each energy level.}
\end{table}

\begin{table}[h]
\centering
\begin{tabular}{c|c|c|c|c}
n&$S^z$&$m_1$&$m_2$&$x^\text{Ising}$\\
\hline \hline 
0&$\pm 1$&0&0&0\\
0&0&0&0&1/2\\
\hline
1&$\pm 1$&1&0&0\\
1&0&1&0&1/2\\
1&0&0&0&3/2\\
\hline 
2&$\pm 2$&0&0&1/2\\
2&$\pm 1$&2&0&0\\
2&$\pm 1$&0&1&0\\
2&$\pm 1$&0&0&2\\
2&0&2&0&1/2\\
2&0&0&1&1/2\\
2&0&1&0&3/2\\
2&0&0&0&5/2\\
\end{tabular}
\caption{Details of $j=1$ conformal tower.}
\end{table}

\begin{table}[h]
\centering
\begin{tabular}{c|c|c|c|}
n&s=0&s=1&s=2\\
\hline 
0&0&1&0\\
1&1&1&0\\
2&1&2&1
\end{tabular}
\caption{j=1 Conformal Tower, labeled by multiplicities of spin multiplets at each energy level.}
\end{table}

\begin{table}[h]
\centering
\begin{tabular}{c|c|c|}
n&s=1/2&s=3/2\\
\hline 
0&1&0\\
1&1&1\\
2&2&2
\end{tabular}
\caption{j=1/2 Conformal Tower, labeled by multiplicities of spin multiplets at each energy level.}
\end{table}

There are selection rules determining which Ising conformal 
towers can combine with boson states of various values of $S^z$.  The 3 conformal towers of $SU(2)_2$ correspond to the following selection rules:
\begin{eqnarray} j&=&0: (S^z=\text{even integer},  I)\  \text{or}\   (S^z=\text{odd integer}, \  \epsilon ) \nonumber \\
j&=&1: (S^z=\text{even integer},  \epsilon )\  \text{or}\   (S^z=\text{odd integer}, \  I ) \nonumber \\
j&=&1/2: (S^z=\text{half-integer}, \sigma ).\label{selrule}
\end{eqnarray}
Using Eq. (\ref{Eb}) and the known multiplicities of the Ising conformal towers we can work out the complete $SU(2)_2$ conformal towers. Knowing the 
$S^z$ quantum numbers allows us to uniquely assign total spin quantum numbers $s$ to multiplets. (Of course, for this to be consistent, the 
number of states at a given level, $n$, for a given $S^z$ must be $\leq$ the number of states at the same level for smaller $S^z$. This turns out to 
always be the case.) Note that half-integer $S^z$ only occurring with $\sigma$ is consistent with the periodicity conditions discussed in  [\onlinecite{J1J2J3_letter}]. These results lead to Tables I-V.

We can  easily read off formulas for the  energy of the lowest state of spin $s$ for any $s$ in each conformal tower.  For the $j=0$ conformal tower, 
the lowest energy state of spin $s$ has energy:
\begin{eqnarray} E&=&{\pi v\over N}\left[-{1\over 16}+{s^2\over 2}\right] ,\ \  (\hbox{s even})\nonumber \\
&=&{\pi v\over N}\left[-{1\over 16}+{s^2\over 2}+{1\over 2}\right] ,\ \  (\text{s odd}).
\end{eqnarray}
For the $j=1$ conformal tower
\begin{eqnarray} E&=&{\pi v\over N}\left[-{1\over 16}+{s^2\over 2}+{1\over 2}\right] ,\ \  (\text{s even})\nonumber \\
&=&{\pi v\over N}\left[-{1\over 16}+{s^2\over 2}\right] ,\ \  (\text{s odd}).
\end{eqnarray}
These results are summarized in Table VI.

\begin{table}[h]
\centering
\begin{tabular}{c|c|c|c|c|c|c|}
s&0&1&2&3&4&5\\
\hline \hline 
$(E-E_0)N/ \pi v$, j=0&0&1&2&5&8&13\\
\hline 
$(E-E_0)N/ \pi v$, j=1&1&0&2&4&8&12
\end{tabular}
\caption{Lowest excitation energy with spin $s$ for both $j=0$ and $j=1$ conformal towers.}
\end{table}

We have checked the validity of the conformal embedding by comparing the characters. The character for a conformal tower is the corresponding partition function:
\begin{equation} \chi_x =q^{-c/12+x}\sum_{n=0}^\infty d_nq^n.\end{equation}
Here $q\equiv e^{-\pi v/(NT)}$, $c$ is the central charge ($3/2$, $1/2$ and $1$ for $SU(2)_2$, Ising and free boson respectively) $x$ labels the conformal tower and the $d_n$'s are the multiplicities. 

For the boson conformal field theory, the characters are given by:
\[
\chi(q)= q^{-\frac{1}{24}}\frac{1}{\varphi(q)}\sum_{S_z} q^{S_z^2}
\]
where $\varphi(q)=\prod_{n=1}^\infty (1-q^n)$ is the Euler function.

If the sum is restricted to even values of $S_z$, this leads to:
\begin{eqnarray*}
\chi^\text{boson}_\text{even}(q)&=& q^{-\frac{1}{24}}\frac{1}{\varphi(q)}\sum_{n} q^{2n^2}\\
&=&q^{-\frac{1}{24}}(1+q+4q^2+5q^3+9q^4+...)
\end{eqnarray*}
while if the sum is restricted to odd values of $S_z$, this leads to: 
\begin{eqnarray*}
\chi^\text{boson}_\text{odd}(q)&=& q^{-\frac{1}{24}}\frac{1}{\varphi(q)}\sum_{n} q^{(2n-1)^2/2}\\
&=& q^{-\frac{1}{24}}q^{\frac{1}{2}}\frac{1}{\varphi(q)}\sum_{n}q^{2n(n-1)}\\
&=& q^{-\frac{1}{24}}q^{\frac{1}{2}}(2+2q+4q^2+6q^3+12q^4+...)
\end{eqnarray*}
Finally, if the sum is restricted to half-integer values of $S_z$, this leads to: 
\begin{eqnarray*}
\chi^\text{boson}_\text{1/2}(q)&=& q^{-\frac{1}{24}}\frac{1}{\varphi(q)}\sum_{n} q^{(n-\frac{1}{2})^2/2}\\
&=&q^{-\frac{1}{24}}q^{\frac{1}{8}}\frac{1}{\varphi(q)}\sum_{n} q^{\frac{n(n-1)}{2}}\\
&=&q^{-\frac{1}{24}}q^{\frac{1}{8}}(2+2q+4q^2+6q^3+12q^4+...)
\end{eqnarray*}

For the Ising conformal field theory, the characters are given by (see [\onlinecite{diFrancesco}] page 242-243):
\begin{eqnarray*}
\chi(q) &=& q^{-\frac{1}{48}}q^{h_{r,s}}\frac{q^{-\frac{1}{48}}q^{-h_{r,s}}}{\varphi(q)}\\
&\times& \sum_{n} \left[q^{\frac{(24 n + 4 r-3s)^2}{48}}-q^{\frac{(24 n + 4 r+3s)^2}{48}}\right]
\end{eqnarray*}
where $h_{r,s}=\frac{(4r-3s)^2-1}{48}$ and where $(r,s)=(1,1)$ for $I$, $(r,s)=(2,1)$ for $\epsilon$, and $(r,s)=(1,2)$ for $\sigma$.

This leads to the following characters for $I$, $\epsilon$ and $\sigma$:
\begin{eqnarray*}
\chi^\text{Ising}_I(q)&=& q^{-\frac{1}{48}}\frac{q^{-\frac{1}{48}}}{\varphi(q)}\sum_{n} \left[q^{\frac{(24 n + 1)^2}{48}}-q^{\frac{(24 n + 7)^2}{48}}\right]\\
&=&q^{-\frac{1}{48}}
(1+q^2+q^3+2q^4+2q^5+...)
\end{eqnarray*}
\begin{eqnarray*}
\chi^\text{Ising}_\epsilon(q)&=& q^{-\frac{1}{48}}q^\frac{1}{2}\frac{q^{-\frac{1}{48}}q^{-\frac{1}{2}}}{\varphi(q)}\sum_{n} \left[q^{\frac{(24 n + 5)^2}{48}}-q^{\frac{(24 n + 11)^2}{48}}\right] \\
&=&q^{-\frac{1}{48}}q^\frac{1}{2}(1+q+q^2+q^3+2q^4+2q^5+...)
\end{eqnarray*}
\begin{eqnarray*}
\chi^\text{Ising}_\sigma(q)&=& q^{-\frac{1}{48}}q^\frac{1}{16}\frac{q^{-\frac{1}{48}}q^{-\frac{1}{16}}}{\varphi(q)}\sum_{n} \left[q^{\frac{(24 n -2 )^2}{48}}-q^{\frac{(24 n + 10)^2}{48}}\right]\\
&=&q^{-\frac{1}{48}}q^\frac{1}{16}(1+q+q^2+2q^3+2q^4+3q^5+...)
\end{eqnarray*}

For the $SU(2)_2$ conformal field theory, the characters for a given $j$ are given by (see [\onlinecite{diFrancesco}] page 585):
\[
\chi(q)= q^{-\frac{1}{16}}q^{\frac{j(j+1)}{4}}\frac{\sum_{n} (2j+1+8n)q^{n(2j+1+4n)}}{\sum_{n} (1+4n)q^{n(1+2n)}}
\]
This leads to the following characters for $j=0,\frac{1}{2}\ \text{and}\ 1$:
\[
\chi^{SU(2)_2}_{j=0}(q)= q^{-\frac{1}{16}}(1+3q+9q^2+15q^3+30 q^4+...)
\]
\[
\chi^{SU(2)_2}_{j=\frac{1}{2}}(q)= q^{\frac{1}{8}}(2+6q+12q^2+26q^3+48 q^4+...)
\]
\[
\chi^{SU(2)_2}_{j=1}(q)= q^{\frac{7}{16}}(3+4q+12q^2+21q^3+43 q^4+...)
\]

By expanding all these characters to  order 100 or higher, we have checked that the following relations hold:
\[
\chi^\text{boson}_\text{even}(q)\chi^\text{Ising}_I(q)+\chi^\text{boson}_\text{odd}(q)\chi^\text{Ising}_\epsilon(q)= \chi^{SU(2)_2}_{j=0}(q)
\]
\[
\chi^\text{boson}_\text{even}(q)\chi^\text{Ising}_\epsilon(q)+\chi^\text{boson}_\text{odd}(q)\chi^\text{Ising}_I(q)= \chi^{SU(2)_2}_{j=1}(q)
\]
\[
\chi^\text{boson}_\text{1/2}(q)\chi^\text{Ising}_\sigma(q)= \chi^{SU(2)_2}_{j=\frac{1}{2}}(q)
\]
corresponding to Eq. (\ref{selrule}).

\subsection{DMRG}

Most of the numerical results in this paper have been obtained with Density Matrix Renormalization Group (DMRG) algorithm. The only exception is the calculation of the Berry phase that has been done on small rings by exact diagonalization. In this section we provide some technical details on the DMRG algorithm we have used.

First of all, we have used the Matrix Product State formulation of DMRG, and therefore the proper name would be variational MPS. The code consists of four parts:

1. Infinite-size DMRG: The system size grows from 2 to $N$ by inserting two-site Matrix Product Operator (MPO) in the middle of the chain and by diagonalizing the corresponding effective Hamiltonian. Everything on the left and on the right of the MPO is effectively described by the left and right environments. The singular value decomposition of the eigenvector produces left and right normalized on-site tensors. These tensors are multiplied with the corresponding environments in order to update them and at the same time to increase the size of each environment by one.
For an odd total number of spins $N$, the same procedure is performed until the system reaches a size of $N-1$, in which case only one tensor is multiplied to the environment. Assuming without loss of generality that the left environment was updated and therefore contains an effective basis for $(N-1)/2$ spins, one can reuse the right environment for $(N-3)/2$ spins and insert the local Hamiltonian for two additional spins to reach a system size with N odd.
In this part of the code, we usually keep 44 singular values. However, close to the critical lines, we increase this number to 66 for systems larger than $N\geq300$ spins. The infinite-size DMRG provides a good starting point for the remaining parts of the code.

2. The warm-up function consists of an incomplete sweep. Sweeping from the middle of the chain to its right end we update local tensors  site-by-site and increase the number of kept states by a factor $1.5$. Sweeping back from the right end to the left one we again increase the number of states by the same factor. Therefore in the end of the warm-up the number of kept states is 100 (or 150 for 66 states in infinite-size DMRG).

3. The 'main body' of the algorithm is sweeping from left to right and back locally updating the tensors. We usually perform 6-7 sweeps for open boundary conditions and up to 16 sweeps for periodic chains. We keep up to 700 singular values for $N<200$ and up to 900 states for larger systems. During the first 6-7 sweeps we increase the number of states linearly up to its maximal value. For periodic chains we continue to jiggle the wave-function by decreasing and increasing the number of states until the convergence is reached. The traditional formulation of the variational MPS imply 'one-site' DMRG, where the effective Hamiltonian diagonalized at each iteration represents a single spins in its left and right environments. Since the dimerized phase has two spins per unit cell, we implement a two-site routine, which turns out to be significantly more stable and to converge faster, despite the obvious growth of complexity by a factor $(2S+1)^2=9$. Roughly speaking, the number of kept states 700 and 900 for two-site DMRG is equivalent to 1210 and 1560 for the one-site routine, although there is no simple one-to-one correspondence.

4. During the 'final sweep' we do not increase the number of states anymore, but at each iteration we measure the set of local observables such as on-site magnetization, nearest-neighbor spin-spin correlations and entanglement entropy. The left and right normalized tensors and vectors of the Schmidt decomposition are stored and used later in order to calculate the observables which involve more than two spins (energy in the middle of the chain, long-range correlations, structure factor, etc.) or to extract the entanglement spectrum.

A significant role in the successful convergence is played by an efficient representation of the Hamiltonian (\ref{eq:H}) in terms MPO. The MPO is a four dimensional tensor with two physical and two auxiliary legs. The complexity of the algorithm is proportional to the dimension of the latest. The straightforward MPO representation of the $J_1-J_2-J_3$ model has a bond dimension $d=17$ ($3+3$ for $J_1$ and $J_2$ terms, 9 for $J_3$ interaction, 1 for unity matrix and 1 for magnetic field or so-called 'full term'). Using the spin commutation relations this number can be reduced to $d=14$. Below we show a different approach that allows to reduce the bond dimension to $d=8$.

The efficient MPO representation naturally appears when the $J_3$-term is rewritten in terms of quadrupolar operators:

\begin{equation}
\sum_{i}\sum_{\alpha,\beta=x,y,z}J_3 S^\alpha_{i-1} Q^{\alpha\beta}_i S^\beta_{i+1},
\end{equation}
where
\begin{equation}
Q^{\alpha\beta}_i=S^\alpha_{i}S^\beta_{i}+S^\beta_{i}S^\alpha_{i}.
\end{equation}
Generally speaking $Q$ is not a traceless tensor, and therefore it is not a quadrupolar operator, but let us keep the $Q$-notation for simplicity. Combining the new expression for the $J_3$ term with the $J_2$ term, one obtains the Hamiltonian in the following form:

\begin{multline}
H=\sum_{i} J_1{\bf S}_i\cdot {\bf S}_{i+1}\\+\sum_{i}\sum_{\alpha,\beta=x,y,z} 
S^\alpha_{i-1} \left(J_2\delta^{\alpha\beta}+J_3 Q^{\alpha\beta}_i \right)S^\beta_{i+1}
\end{multline}

The sum in brackets can be written in matrix form as:

\begin{equation}
 \begin{pmatrix}
   J_2I+J_3Q^{xx} & J_3Q^{xy} & J_3Q^{xz} \\
   J_3Q^{xy}& J_2I+J_3Q^{yy} & J_3Q^{yz} \\
   J_3Q^{xz}&J_3Q^{yz} & J_2I+J_3Q^{zz}
 \end{pmatrix},
\end{equation}
where $I$ is a $d\times d$ unity matrix, with $d=2S+1$. In terms of rescaled lowering and raising operators $S_i^\pm=(S_i^x\pm iS^y_i)/\sqrt{2}$, the MPO tensor reads:

\begin{widetext}

\begin{equation}
\label{eq:mpo}
H_{i} = 
 \begin{pmatrix}
   I & & & &  & & & \\
   S^-_i & & & &  & & & \\
   S^+_i & & & &  & & & \\
   S^z_i & & & &  & & & \\
   & J_2I+J_3Q^{+-} & J_3Q^{--} & J_3Q^{-z}&  & & & \\
   & J_3Q^{++}& J_2I+J_3Q^{+-} & J_3Q^{+z}&  & & & \\
   & J_3Q^{+z}&J_3Q^{-z} & J_2I+J_3Q^{zz}&  & & & \\
   hS^z_i& J_1S^+_i&J_1S^-_i & J_1S^z_i&S^+_i&S^-_i&S^z_i & I
 \end{pmatrix},
\end{equation}

\end{widetext}
where blank spaces correspond to zero entries. The MPO Hamiltonians for the first and the last sites are given by the last row and the first column of tensor (\ref{eq:mpo}) respectively.

\begin{figure}[h!]
\includegraphics[width=0.49\textwidth]{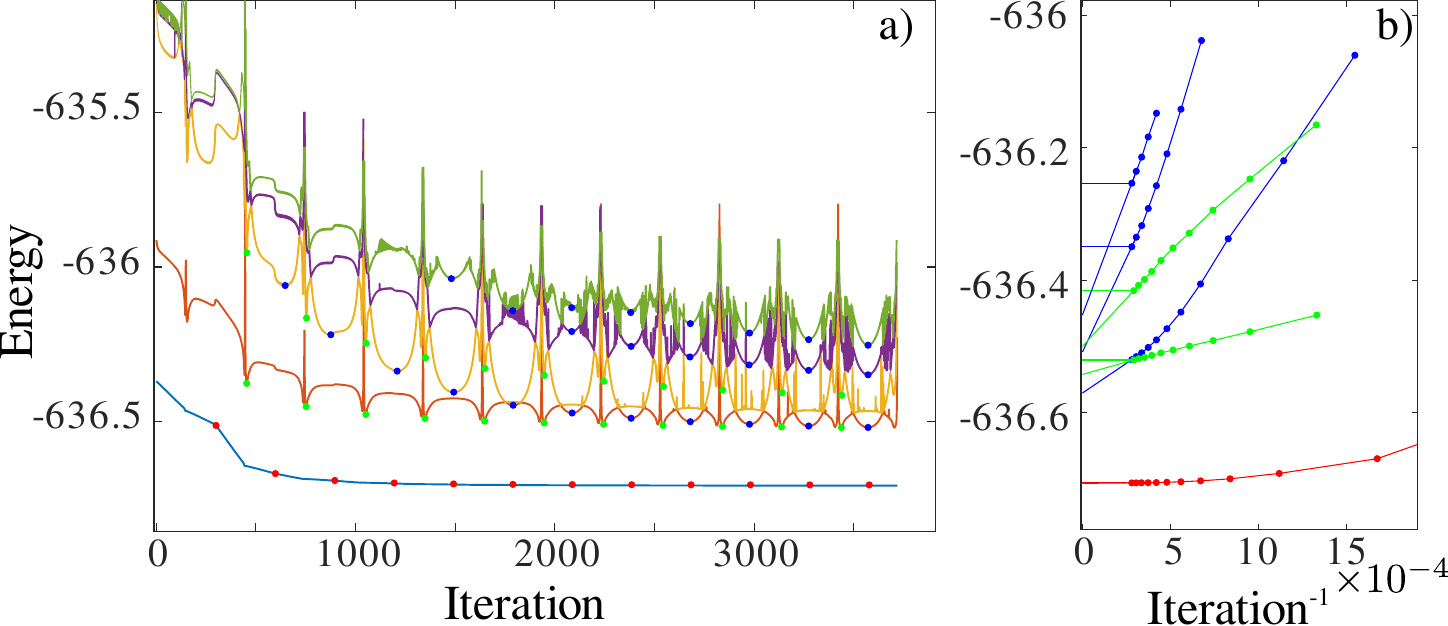}
\caption{(Color online) a) Energy of the ground state and of the first four low-lying excited states as a function of the number of DMRG iterations. The ground-state energy, bulk and edge excitation are marked with red, blue and green dots respectively. b) Energy scaling with the number of iterations. The values of the energies are taken at the points marked on the left panel. The lower bound estimates are linear fits of the last two available points. The upper bounds are the values of the last available points.}
\label{fig:dmrg_edge}
\end{figure}

The investigation of the low-lying spectra along the Ising critical line requires to be able to access the energy of several low-lying states within one symmetry sector. For finite-values of the nearest-neighbor coupling $J_1$ the triplet excitation is a bulk excitation and it is above a few low lying singlet excitations for systems with $N>30$ \cite{J1J2J3_letter}. The picture is more complicated when the $J_1$ coupling is absent. The bonds close to each edge can be excited to a triplet state with a very low energy. The singlet bulk excitations are below the edge excitations for very large system sizes $N>300$, see Fig.\ref{fig:j2j3_ct}. When bulk and edge excitations are close enough, they can be distinguished by looking at the excitation energy as a function of iterations. The energies obtained by diagonalizing the effective Hamiltonian with an MPO located in the middle of the chain are related to the bulk excitations, while minima in energies around the end of each half-sweep correspond to edge excitations. In Fig.\ref{fig:dmrg_edge} we provide an example for the $J_2-J_3$ chain with $N=300$ spins.

\section{Phase diagram}
\label{sec:methods}
\subsection{Dimerization}

\begin{figure}[h!]
\includegraphics[width=0.4\textwidth]{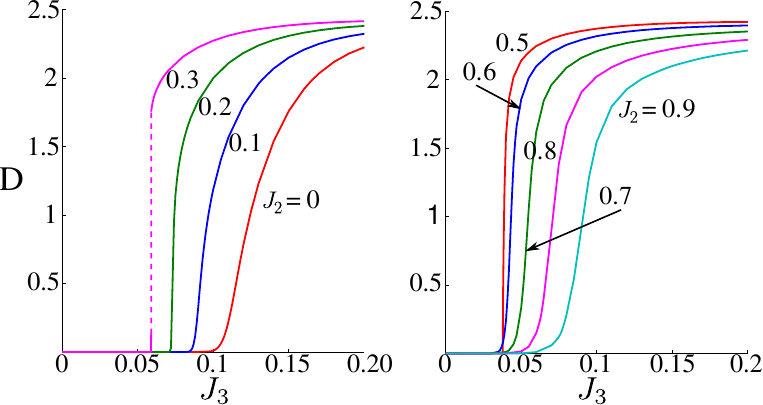}
\caption{(Color online) Finite-size $(N=150)$ dimerization as a function of $J_3$ for $0\leq J_2\leq 0.3$ (left panel) and $0.5\leq J_2\leq 0.9$ (right panel). The dashed line indicates a first order phase transition.}
\label{fig:Dimerization}
\end{figure}

The natural order parameter to identify the dimerized phase is the dimerization parameter defined by $D=|\langle{\bf S}_i\cdot {\bf S}_{i+1}-{\bf S}_i\cdot{\bf S}_{i-1}\rangle|$ where $(i,i+1)$ is the central bond of an open chain.  
Fig.\,\ref{fig:Dimerization} shows numerical results for the dimerization of a chain with $N=150$ sites as a function of  $J_3$ obtained by variational MPS.

In large systems ($N=120,150$) and close to the first order phase transition the variational MPS algorithm suffers from a kind of hysteresis: the algorithm converges to the first excited state instead of the ground state. This results in an unphysical jump in the energy curve and in an abrupt change of dimerization before the actual phase transition. These results were discarded when discussing the nature of the phase transition, and Fig.\,\ref{fig:Dimerization} presents only dimerization curves for which the finite-size energy is continuous.

\begin{figure}[t]
\includegraphics[width=0.49\textwidth]{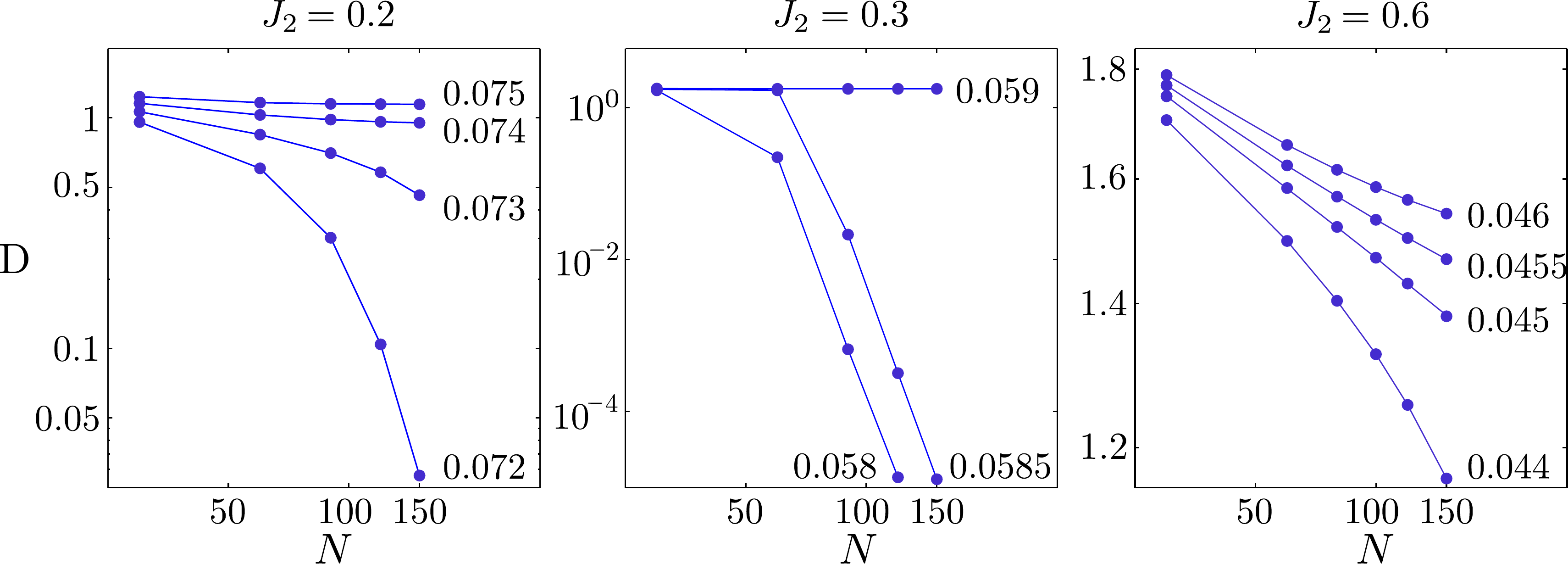}
\caption{(Color online) Finite-size scaling of the dimerization parameter for $J_2=0.2,\,0.3$ and $0.6$. The value of $J_3$ is attached to each curve. The phase transition is continuous at $J_2=0.2$ and $0.6$. The abrupt change of scaling at $J_2=0.3$ indicates a first order phase transition.}
\label{fig:DimeriScaling}
\end{figure}

In order to determine the boundary of the dimerized phase in the thermodynamic limit we have performed a finite-size extrapolation for chains with $N=30,\ 60,\ 90,\ 120$ and $150$ sites. A chain is in the dimerized phase if the dimerization stays finite for $N\rightarrow\infty$, which we associate with a convex curve in a $\log-\log$ plot. By contrast, a concave scaling curve leads to a vanishing dimerization in the thermodynamic limit and therefore means that the system is in the Haldane or NNN-Haldane phase. The phase transition then corresponds to a straight line in the scaling. Some examples of finite-size scaling are shown in Fig.\,\ref{fig:DimeriScaling}. A smooth change of the scaling curvature implies that the dimerization curve is continuous in the thermodynamic limit Fig.\ref{fig:DimeriScaling} a,c), while a first order phase transition with a finite jump in the dimerization curve leads to an abrupt change from concave to convex scaling at the critical point Fig.\ref{fig:DimeriScaling}  b).

The investigation of dimerization has led to a precise determination of the transition line, and of the nature of the phase transition (continuous or first order) except in the 
vicinity of the end point of the continuous transition between the Haldane phase and the dimerized phase (see section E).

\subsection{Ground-state energy}

\begin{figure}[t]
\includegraphics[width=0.47\textwidth]{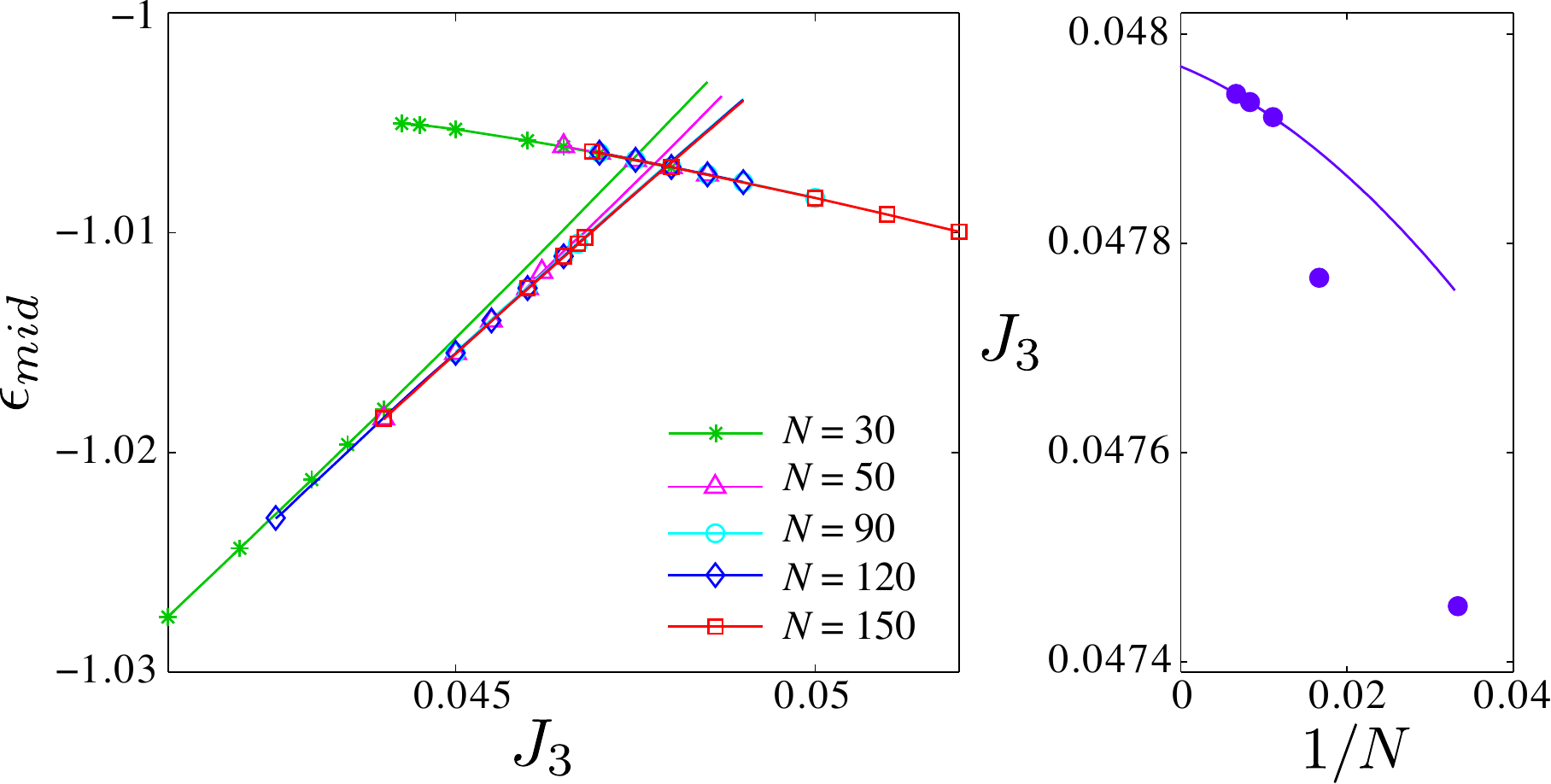}
\caption{(Color online) a) Energy of the central bonds for $J_2=0.4$ as a function of $J_3$ for finite-size chains with $N=30,60,90,120$ and $150$ sites. Solid lines are polynomial fits. The kink was created by letting the two fits cross. b) Position of the kink in $\epsilon_{mid}$ as a function of size. The fitting curve is a quadratic function in $1/N$.}
\label{fig:energy04}
\end{figure}

In view of the hysteretic behavior of the system, and to complete the phase diagram in regions where the ground-state energy appears to be discontinuous, 
we have carefully investigated  the behavior of the energy in the vicinity of the transition lines.
An estimate of the ground-state energy in the thermodynamic limit is provided by the value of the energy of the central bonds: 
$$\epsilon_{mid}=\epsilon_1+\epsilon_2+\epsilon_3,$$
where
$$\epsilon_1=\frac{J_1}{2}\langle{\bf S}_{i-1}\cdot {\bf S}_i+{\bf S}_i\cdot{\bf S}_{i+1}\rangle,$$
$$\epsilon_2=J_2\langle{\bf S}_{i-1}\cdot {\bf S}_{i+1}\rangle,$$
$$\epsilon_3=J_3\langle({\bf S}_{i-1}\cdot {\bf S}_i)({\bf S}_i\cdot {\bf S}_{i+1})+{\mathrm H.c.}\rangle,$$
and where $(i,i+1)$ is the central bond.
The dependence on $J_3$ of $\epsilon_{mid}$ for chains with $N=30,\ 60,\ 90,\ 120$ and $150$ sites for $J_2=0.4$ is presented in Fig.\,\ref{fig:energy04}a). The energy curves are discontinuous due to the edge effects and due to hysteresis of the variational MPS algorithm for $N=120,150$. In order to determine as precisely as possible the location of the first order phase transition in the thermodynamic limit, we have extrapolated the lines until they cross. 
Then, a finite-size scaling of the position of the kink is presented in Fig.\,\ref{fig:energy04}b). The result is in very good agreement with the position determined form $\epsilon_{mid}$, which we take as our estimate of the location of the first-order transition.

\begin{figure}[t]
\includegraphics[width=0.47\textwidth]{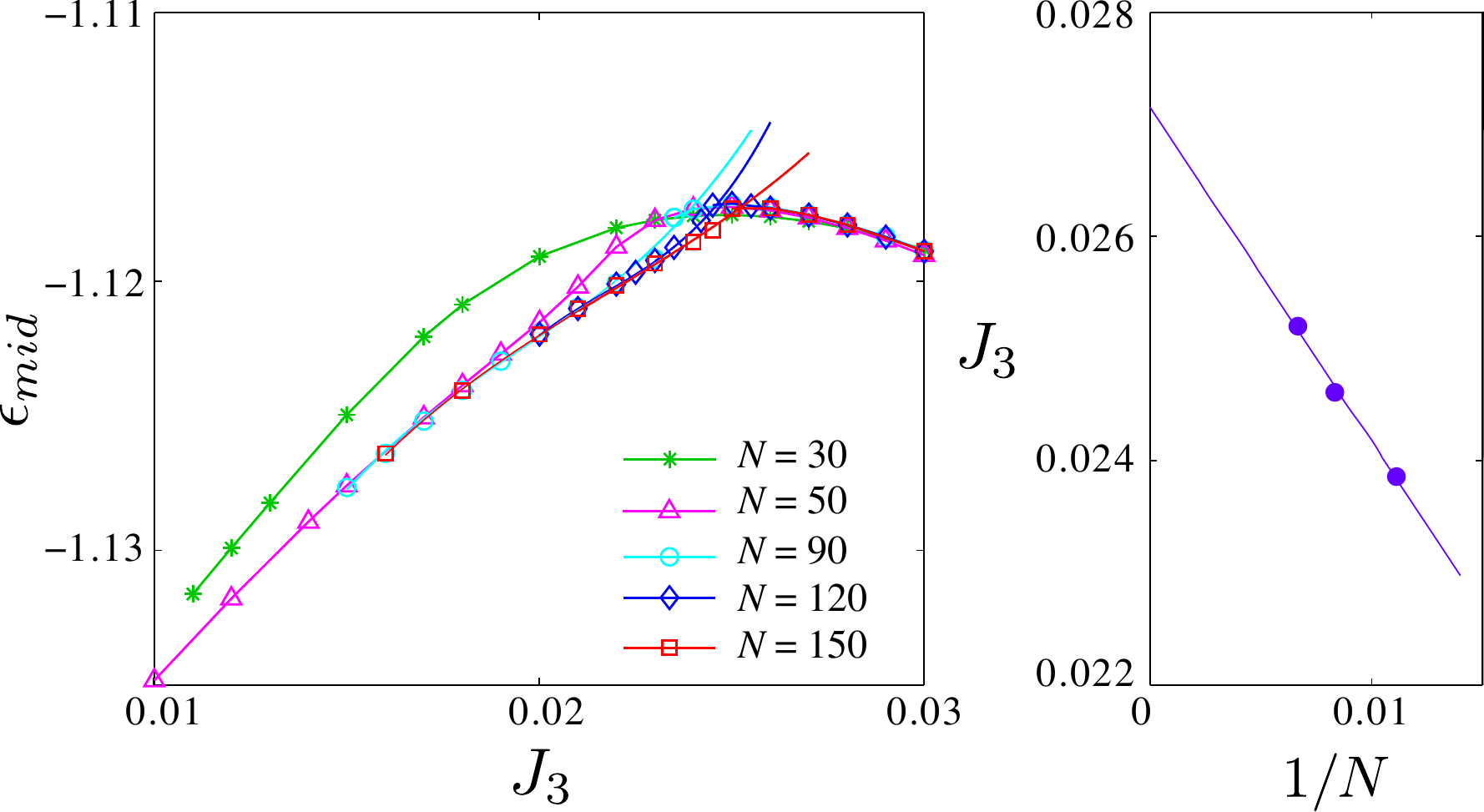}
\caption{(Color online) Same as Fig.\,\ref{fig:energy04} for $J_2=0.6$}
\label{fig:energy06}
\end{figure}

For $0.25\leq J_2\leq 0.45$, the ground-state energy and the dimerization parameter lead to the same estimate for the location of the phase transition. For larger next-nearest-neighbor coupling, the kink disappears for small clusters but it is still present in large chains (see Fig.\,\ref{fig:energy06}). The phase transition line continues towards small $J_3$ and end up at $J_3=0$ and $J_2=0.75$, close to the value $0.0744(4)$ obtained by Kolezhuk et al. \cite{kolezhuk_prl,kolezhuk_prb}. 

\begin{figure}[t]
\includegraphics[width=0.47\textwidth]{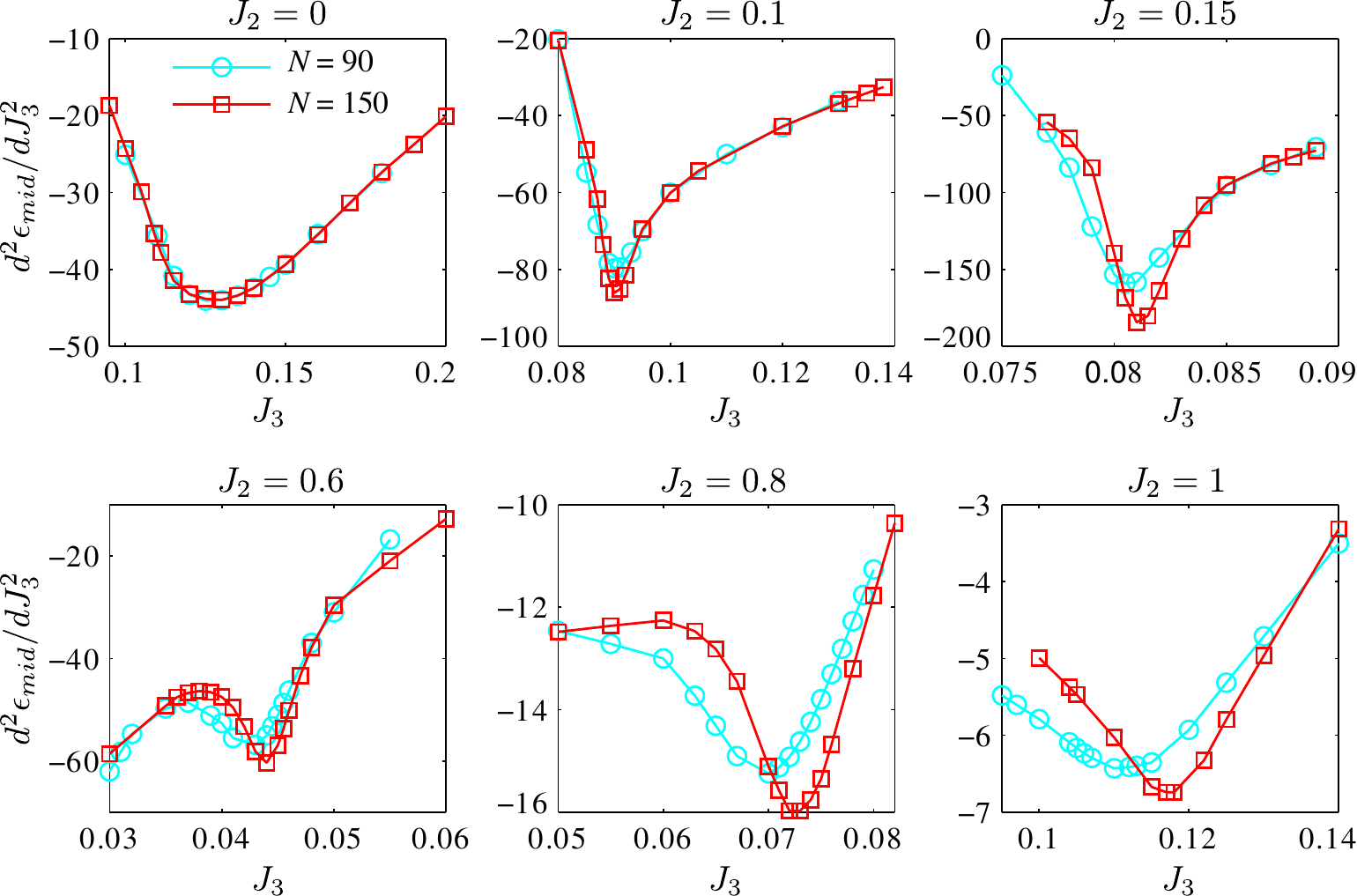}
\caption{(Color online) Upper panels: second derivative of $\epsilon_{mid}$ with respect to $J_3$ for $N=90,150$ and $J_2=0,0.1,0.15$ across continuous phase transition between Haldane and dimerized phases. Lower pannels: second derivative of $\epsilon_{mid}$ for $N=90,150$ and $J_2=0.6,0.8,1$ across the transition line between NNN-Haldane and dimerized phases}
\label{fig:d2energy}
\end{figure}

In order to confirm the location of the continuous phase transition deduced from the dimerization parameter, we have calculated the second derivative of $\epsilon_{mid}$ with respect to $J_3$. Examples for $N=90$ and $150$ are shown in Fig.\,\ref{fig:d2energy}. A kink in the energy implies a divergence of its second derivative. Besides divergences, one can see the appearance of pronounced minima, which agree with the continuous finite-size phase transitions found with the dimerization parameter. For the first order transition between the Haldane phase and the NNN-Haldane phase, a kink is visible in the energy only for non-zero $J_3$. In order to extract the phase boundary at $J_3=0$, and although the phase transition is believed to be first order at this point, we have looked at the minimum of the second derivative of the energy. The finite-size effect slightly increase with increasing $J_2$. The positions of the minima are in good agreement with the phase boundaries found with the dimerization parameter for $J_2\leq0.2$ and $J_2\geq0.5$ (see scaling comparison on Fig.\,\ref{fig:CompareScaling}).

\subsection{Entanglement spectrum}

As mentioned in the introduction, the Haldane phase of the spin-$1$ chain is an example of a symmetry protected topological phase in one dimension\cite{gu}. It is distinct from the topologically trivial NNN-Haldane and dimerized phases, and it can be characterized by the finite value of
the string order parameter, a criterion already used for the $J_1 - J_2$ model\cite{kolezhuk_prl}.
More recently, it has been proposed to characterize topological phases by their entanglement spectrum, obtained by dividing the system into two parts, tracing out one of them, and diagonalizing the reduced density matrix of the remaining part\cite{PhysRevLett.101.010504,PhysRevLett.96.110405,PhysRevLett.96.110404}. This creates artificial edges without breaking the inversion symmetry.

In the present case, a system with open ends may be partitioned across a certain bond and the wave function can be then Schmidt decomposed as: 
\begin{equation}
|\Psi\rangle=\sum_\alpha\lambda_\alpha |L_\alpha\rangle |R_\alpha\rangle,
\end{equation}
where $|L_\alpha\rangle$ and $|R_\alpha\rangle$ are orthonormal basis vectors of the left and right parts. In variational MPS, the Schmidt values 
$\lambda_\alpha$ are obtained naturally at each iteration.
Now, the multiplicity of the Schmidt values is related to the number of edge states that appear due to partitioning: 
Any topologically non-trivial phase is characterized by at least two-fold degeneracy. Pollmann et al.\cite{pollmann} have shown that the Haldane phase of $S=1$ chains is characterized by a twofold degeneracy of the entanglement spectrum.

\begin{figure}[t]
\includegraphics[width=0.45\textwidth]{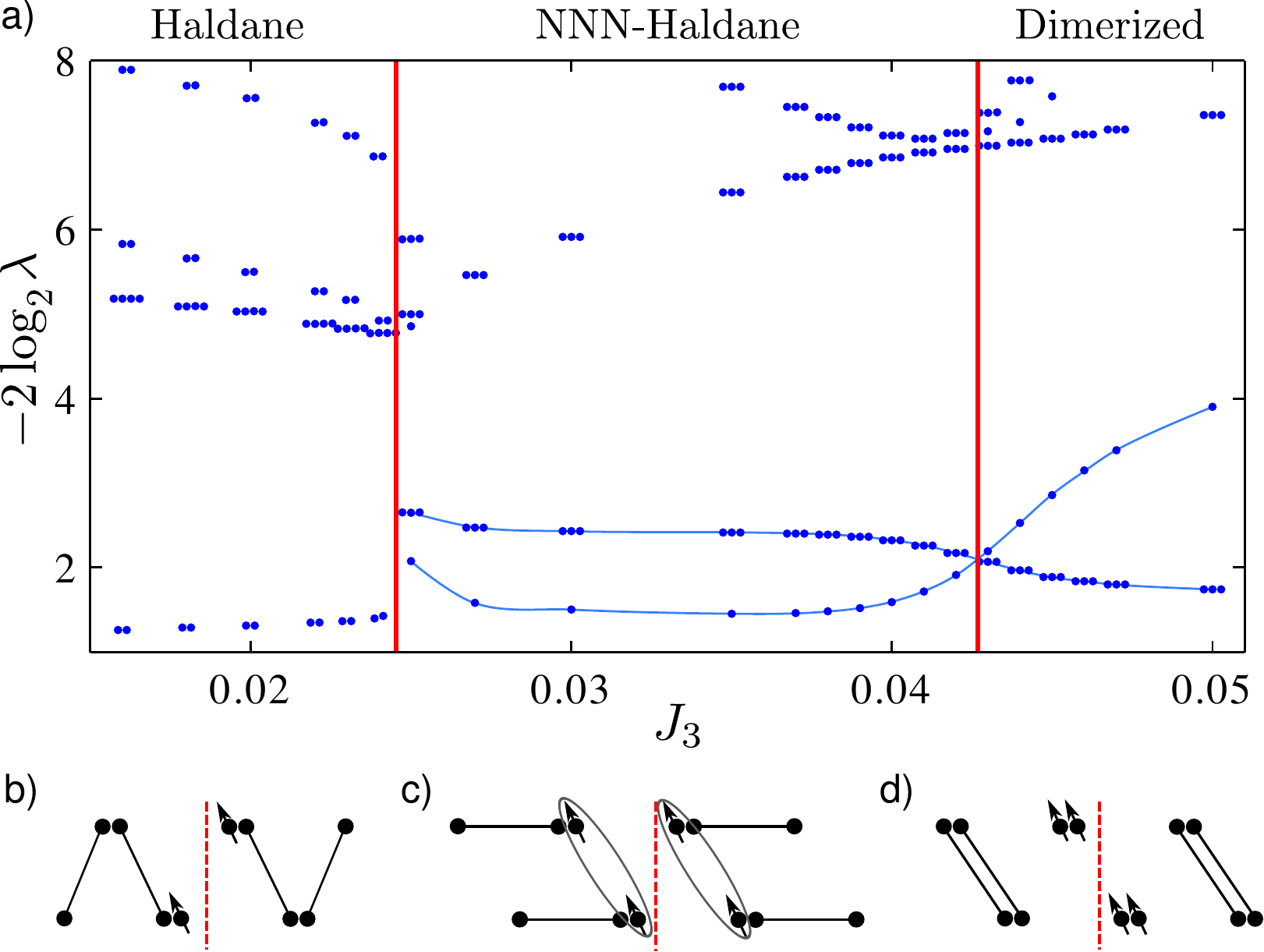}
\caption{(Color online) a) Entanglement spectrum for an open chain with $N=150$ sites as a function of $J_3$ (only the lower part of the spectrum is shown). The dots show the multiplicity of the Schmidt values. The plot for $J_2=0.6$ is shown here as an example. b),c),d) VBS sketches of the artificial edges created by the bipartition of the chain in Haldane, NNN-Haldane and dimerized phase respectively. b) The black arrows at each edge stand for two free spins $1/2$, which form a singlet, leading to a twofold degenerate entanglement spectrum. c) The two spin-$1/2$ created at each edge couple with each other, which is represented as a gray ellipse. There are no edge states, and the entanglement spectrum is non-degenerate. d) The edge spins are spins 1 and form a singlet, which leads to a three-fold degenerate entanglement spectrum.}
\label{fig:EntSpec}
\end{figure}

An example of finite-size entanglement spectrum containing all three phases is shown in Fig.\,\ref{fig:EntSpec}. Three VBS (valence bond solid) sketches are attached in order to show how edge states are formed in each phase. In complete agreement with previous works, the entanglement spectrum in the Haldane phase is twofold degenerate, the edge states being spins 1/2. By contrast, the entanglement spectrum is non-degenerate in the NNN-Haldane phase because there are no edge states. For the dimerized phase, it depends where the system is cut. For a system with open boundary conditions and an even number of sites, the ground state is non degenerate and consists of alternating strong and weak bonds. If the system is cut in the middle of a weak bond, no edge states appear, and the entanglement spectrum is non degenerate. However, if the system is cut on a strong bond, i.e. on 
a bond which is essentially a singlet made of two spins 1, as done in Fig.~\ref{fig:EntSpec}, the entanglement spectrum is threefold degenerate because spin-1 edge states are created, and the NNN-Haldane phase can be distinguished from the dimerized phase. In small systems an intermediate phase with a three fold degenerate entanglement spectrum and a low-lying non-degenerate level appears between Haldane and NNN-Haldane phases. This phase disappears for
larger system sizes and is a thus a finite-size effect.

\begin{figure}[t]
\includegraphics[width=0.48\textwidth]{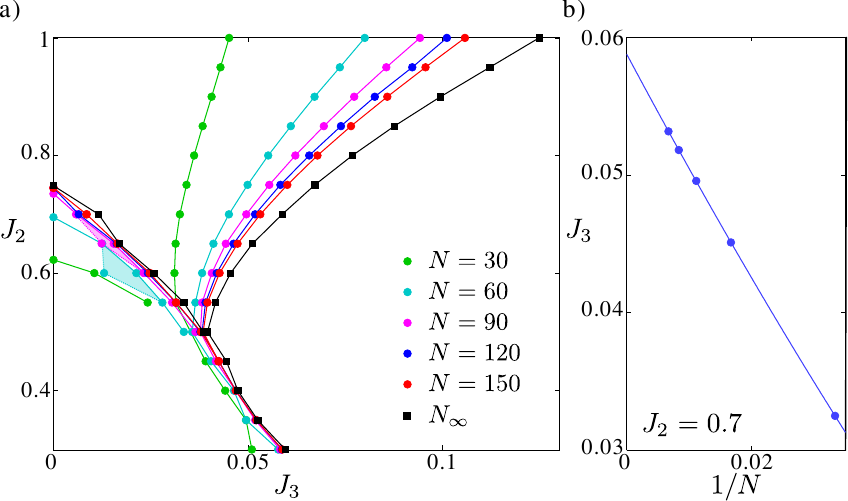}
\caption{(Color online) a) Phase boundaries deduced from the entanglement spectrum for chains with $N=30$, $60$, $90$, $120$, and $150$ sites, and after finite-size scaling ($N_\infty$). Shaded area:  intermediate phase for $N=60$ and $90$. b) Example of finite-size scaling for $J_2=0.7$ with a quadratic fit.}
\label{fig:EntSpecScaling}
\end{figure}

The resulting phase diagram is shown in Fig.~\ref{fig:EntSpecScaling}. It is consistent with other approaches, but finite-size effects are strong, especially
for the transition between the NNN-Haldane phase and the dimerized phase. 

\subsection{Berry Phase}

Another powerful tool to characterize topologically non-trivial phases is the Berry phase\cite{berry}, that can be defined for any Hamiltonian $H(\phi)$ which depends periodically on a parameter $\phi$. If $| GS(\phi)\rangle $ is a single-valued ground state of $H(\phi)$, the Berry connection is given by
$A(\phi)=\langle GS(\phi)|\partial_{\phi}|GS(\phi)\rangle$, and the Berry phase is the integration of the Berry connection over a loop:

$$i\gamma=\oint A(\phi)d\phi$$

It was proposed  by Hatsugai et al.\cite{hatsugai} to use the angle $\phi$ of the twist of the  transverse component of the spin-spin interaction on a given bond $(i,j)$

$$S^+_iS^-_j+S^-_iS^+_j \rightarrow e^{i\phi}S^+_iS^-_j+e^{-i\phi}S^-_iS^+_j.$$

Then the number of valence bond singlets $B_{ij}$ on the bond $(i,j)$ is related to the Berry phase by: 
$$\gamma=B_{ij}\cdot\pi, \ \ \ \ \ \ \mathrm{mod} (2\pi).$$ In other words, the Berry phase gives access to the parity
of the number of valence bond singlets on a given bond.

\begin{figure}[t]
\includegraphics[width=0.47\textwidth]{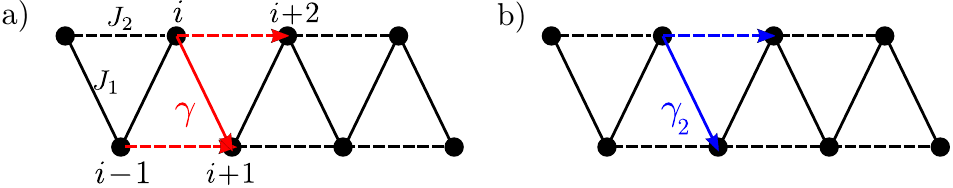}
\caption{(Color online) a) Berry phase applied on three bonds to be consistent with anti-periodic boundary conditions for the transverse component of the spin-spin interaction at $\phi=\pi$. This Berry phase  $\gamma=\pi$ in the Haldane phase and $\gamma=0$ otherwise. b) Berry phase applied on two bonds to distinguish the dimerized phase ($\gamma_2=0$) from the Haldane and NNN-Haldane phases ($\gamma_2=\pi$).}
\label{fig:berry1}
\end{figure}

Previous studies of the Berry phase in spin systems have demonstrated that topological phase transitions can be reliably captured  when the applied twist at $\phi=\pi$ is equivalent to anti-periodic boundary conditions for the transverse component of the interaction.
To fulfill this requirement, three bonds must be simultaneously twisted as shown in Fig.\,\ref{fig:berry1}a. The twist applied on a bond $(i,i+1)$ implies that the transverse component of the interaction in the initial Hamiltonian is changed in all terms where the term ${\bf S}_i\cdot{\bf S}_{i+1}$ appears, i.e. both in  the $J_1$ and $J_3$ terms. The twist of an $(i,i+2)$ bond changes only the $J_2$ term.

In the dimerized phase, there is no singlet on next-nearest neighbor bonds $(i-1,i+1)$ and $(i,i+2)$, while bonds $(i,i+1)$ have either zero or two singlets. So the Berry phase $\gamma$, which is defined only up to $2\pi$, is equal to zero. In the NNN-Haldane phase, the bonds $(i-1,i+1)$ and $(i,i+2)$ contain one spin-$1/2$ singlet each and there is no singlet on the link $(i,i+1)$, so that $\gamma=0$ as in the previous case. By contrast, the Berry phase is equal to 
$\pi$ in the Haldane phase, in which there is one VBS singlet on the bond $(i,i+1)$ and no VBS singlet on the bonds $(i-1,i+1)$ and $(i,i+2)$. 

\begin{figure}[t]
\includegraphics[width=0.49\textwidth]{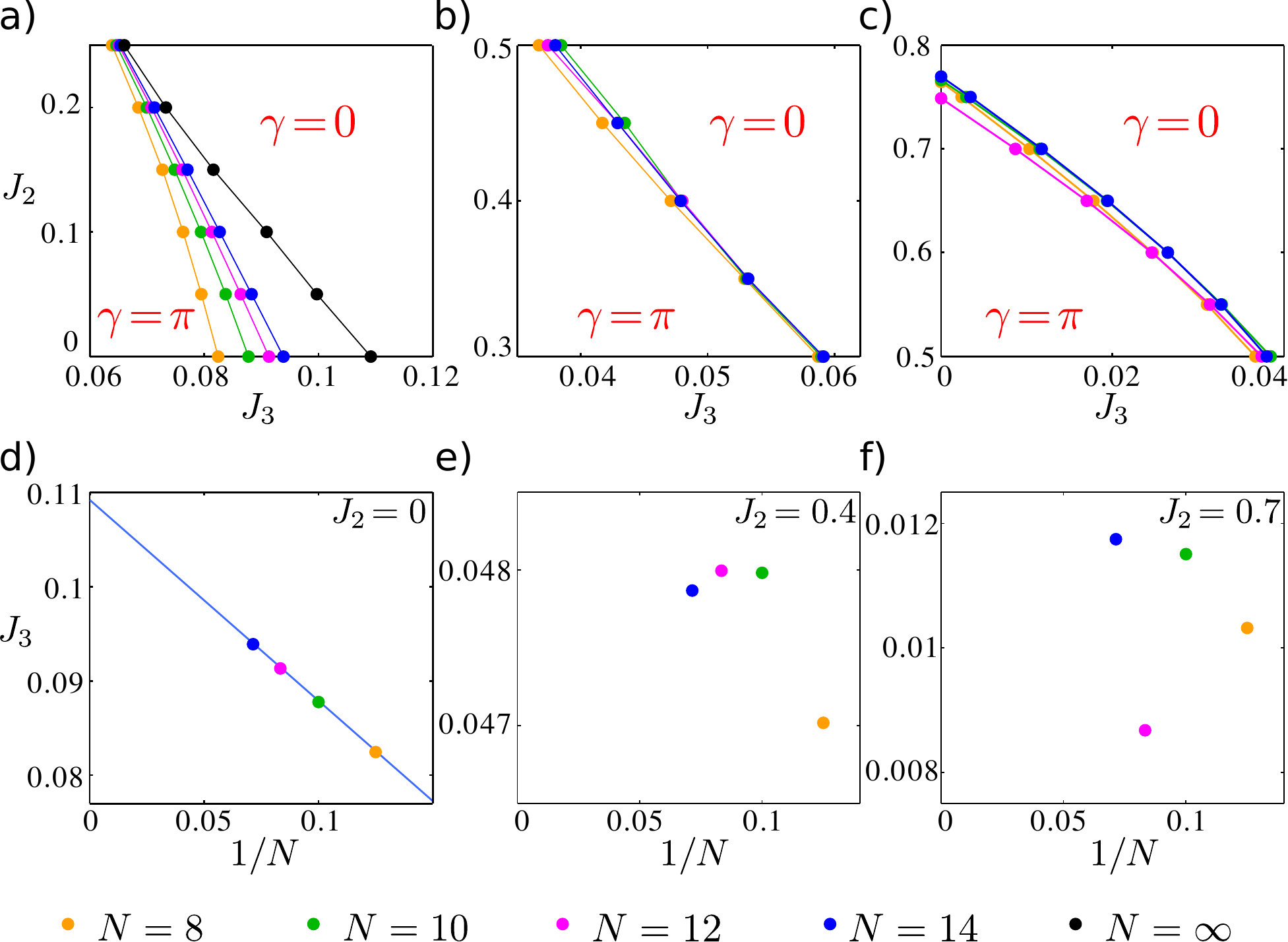}
\caption{(Color online) Results of the Berry phase calculation in a periodic chain with $N=8,10,12,14$. Upper panels: Finite size phase transitions captured by the Berry phase for $J_2$ in the range a) $0\leq J_2\leq 0.25$, b) $0.3\leq J_2\leq 0.5$, and c) $0.5\leq J_2\leq 0.78$. The results of finite-size extrapolation are shown in a) as a black line. Lower panels: Size dependence of the boundaries deduced from Berry phase. d) Finite-size scaling for $J_2=0$ performed with a cubic fit in $1/N$. e) and f) Examples of strong finite-size effects that do not allow one to make a finite-size extrapolation.}
\label{fig:berry}
\end{figure}

We have calculated the Berry phase $\gamma$ for chains with periodic boundary conditions using exact diagonalizations. The results for finite sizes are presented in Fig.\,\ref{fig:berry}a-c, and the finite-size scaling based on chains of length $N=8,10,12,14$ sites in the interval $0\leq J_3\leq 0.25$ is shown in Fig.\,\ref{fig:berry}d). The results from the finite-size extrapolation are also included in Fig.\,\ref{fig:berry}a). Systems close to the first order phase transition have strong finite-size effects, and no meaningful extrapolation could be performed with only four points. There is also a clear indication of an even-odd effect: the scaling for $N=8,12,...,4k$ is different from the one for $N=10,14,...,2(2k+1)$. For $J_2\geq 0.3$, the results for $N=12$ (the largest accessible chain with an even number of spin pairs) is taken as the Berry phase estimate of the phase boundary. Quite remarkably, the finite-size results for such small chains are already very close to the phase boundaries obtained in the thermodynamic limit with other techniques (see Fig.\ref{fig:CompareScaling} and \ref{fig:PDcom}).

\begin{figure}[t]
\includegraphics[width=0.4\textwidth]{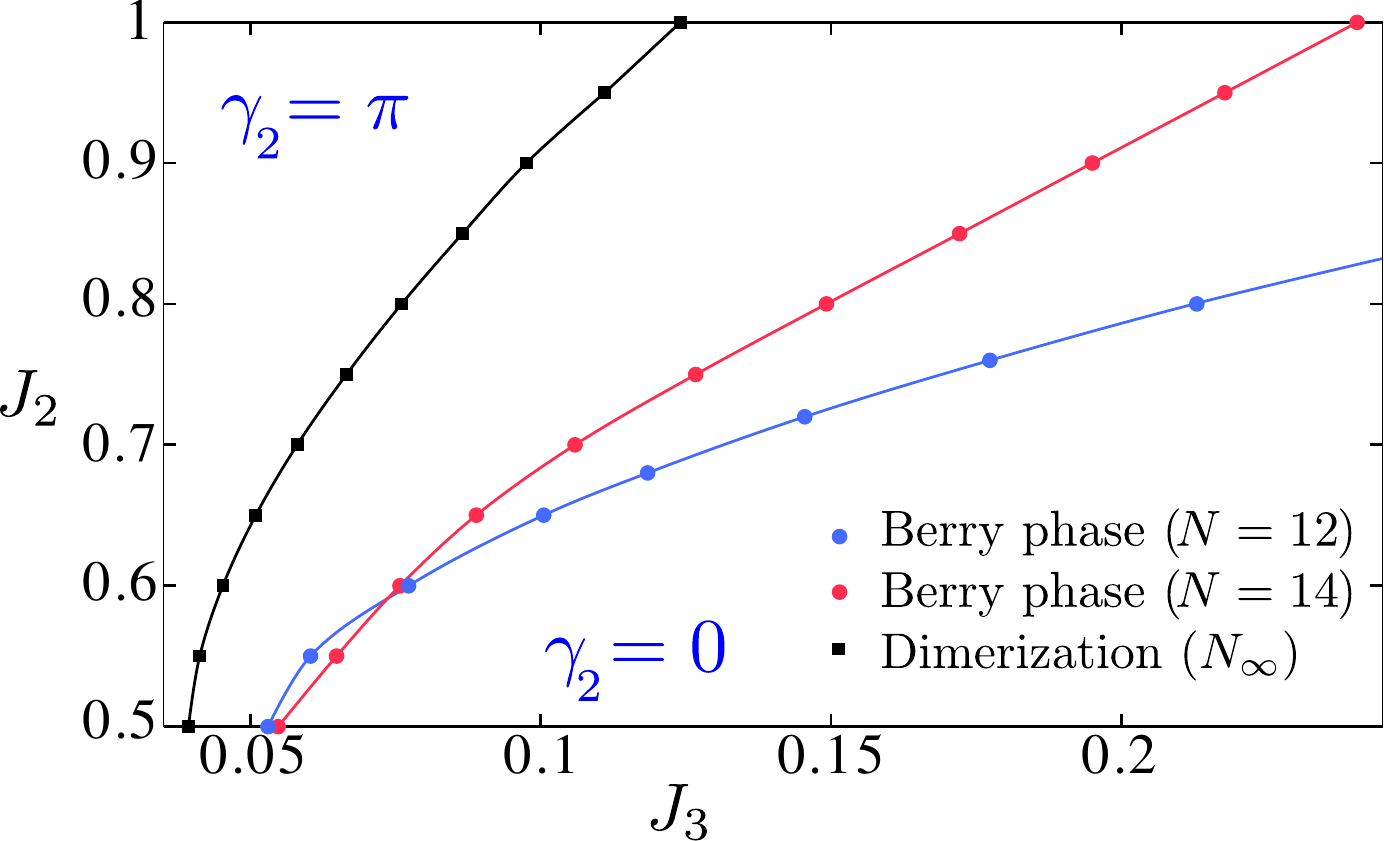}
\caption{(Color online) Phase transition obtained by the $\gamma_2$ Berry phase for periodic chains with $N=12$ (blue circles) and $N=14$ (red circles) sites. The phase transition, obtained with finite-size scaling of the dimerization parameter (black squares) is shown as a reference.}
\label{fig:berry2}
\end{figure}

When the twist used to define the Berry phase does not correspond to anti-periodic boundary conditions at $\phi=\pi$, the Berry phase can still reflect some local properties of the system and capture phase transitions. To distinguish the dimerized phase from the NNN-Haldane phase, we propose to define the Berry phase by twisting two links as shown in Fig.\,\ref{fig:berry1}b. Similarly to what was done for the three-bond Berry phase, we apply the twist on two bonds $(i,i+1)$ and $(i,i+2)$ simultaneously. In the Haldane phase there is only one VBS singlet on the bond $(i,i+1)$, in the NNN-Haldane phase one VBS singlet on the $(i,i+2)$ bond, and in both cases $\gamma_2=\pi$. In the dimerized phase the bond $(i,i+1)$ contains either zero or two singlets while the $(i,i+2)$ bonds have no singlets, and the Berry phase $\gamma_2=0$. The finite-size results for $N=12$ and $N=14$ are shown in Fig.\,\ref{fig:berry2}. Qualitatively, this Berry phase gives the same phase boundaries as the dimerization parameter, which is also shown as a reference line. There is a strong finite-size effect however, and the extrapolation to the thermodynamic limit requires bigger system sizes.

\subsection{Comparison}

\begin{figure}[t]
\includegraphics[width=0.45\textwidth]{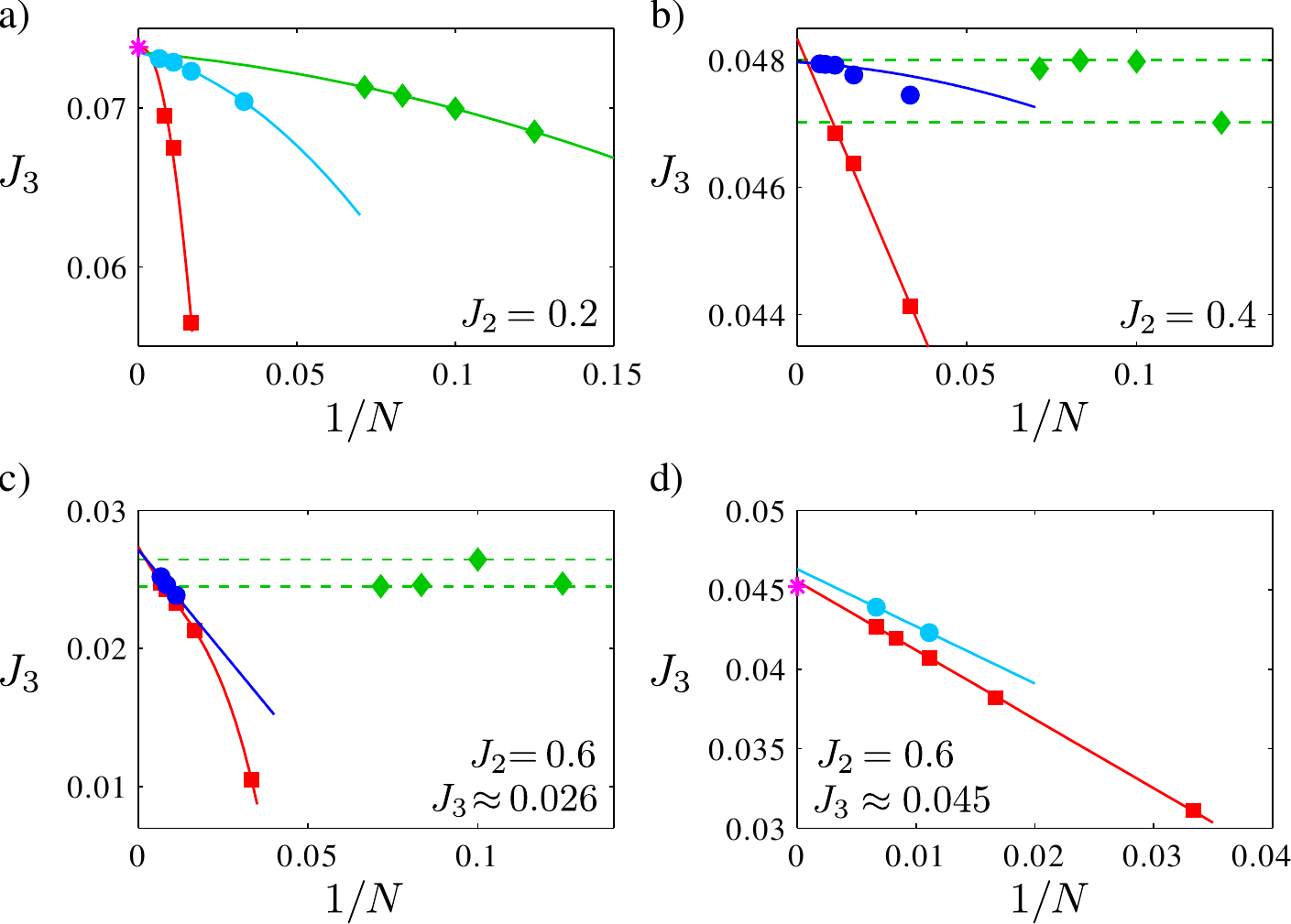}
\caption{(Color online) Comparison of finite-size scaling for a) $J_2=0.2$, b) $J_2=0.4$, c) $J_2=0.6$ and $J_3\simeq 0.026$, d) $J_2=0.6$ and $J_3\simeq 0.045$. Finite-size results for: entanglement spectra (red squares), the Berry phase (green diamonds), kink in the energy of the central bond $\epsilon_{mid}$ (blue circles), minimum in the second derivative of  $\epsilon_{mid}$ (cyan circles). Results from finite-size scaling of the dimerization parameter (magenta stars). All fitting curves are polynomial in $1/N$. Dashed green lines shows the interval between the smallest and the biggest values deduced from the Berry phase.}
\label{fig:CompareScaling}
\end{figure}

\begin{figure}[h!]
\includegraphics[width=0.4\textwidth]{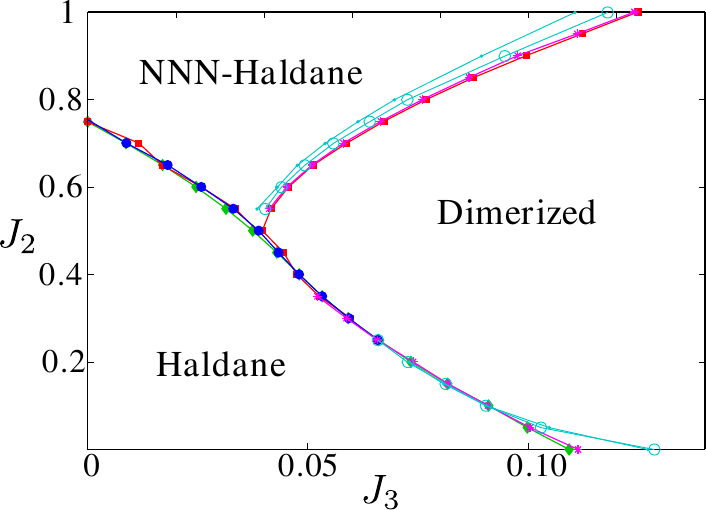}
\caption{(Color online) Comparative phase diagram obtained by dimerization parameter (magenta stars), kink in the energy of the central bond $\epsilon_{mid}$ (blue circles), $\epsilon_{mid}(N=150)$ and $\epsilon_{mid}(N=90)$ (cyan open circles and dots), entanglement spectra (red squares), and Berry phase (green diamonds)}
\label{fig:PDcom}
\end{figure}

To show that all approaches presented above capture essentially the same phase diagram, we provide examples of comparative finite-size scaling (Fig.\,\ref{fig:CompareScaling}) and phase diagrams obtained with different criteria (Fig.\,\ref{fig:PDcom}).

We compare the phase boundaries deduced from the dimerization, energy, entanglement spectra and Berry phase (Fig.\,\ref{fig:PDcom}). For $J_2=0$ the second derivative in the energy gives a phase boundary different from the one obtained with the dimerization parameter. Except for this point, the two boundaries are in rather good agreement. They also agree with the 'entanglement boundary' between the NNN-Haldane phase and the dimerized phase.
The first order phase transition from Haldane to the dimerized phase is well located by all methods. The most reliable phase boundary between Haldane and NNN-Haldane phases is obtained by the kink in the energy of the central bond. Since on the one hand, the kink in the $\epsilon_{mid}$ for large $J_3$ has vanishing finite-size effect (see Fig.\,\ref{fig:energy04}b) and on the other hand the kink for small $J_3$ appears only in large systems, we determined the boundary of the phases with the energy of the central bonds of the largest cluster to which we have access $\epsilon_{mid}(N=150)$.
We cannot see a kink for $J_3=0$ and to locate the phase transition on the $J_2$ axis we have used the minimum in the second derivative of the energy with respect to $J_2$.
 The error in the 'entanglement boundary' is due to the abrupt change of the degeneracy from two in the Haldane phase to one in the NNN-Haldane or three in the dimerized. The finite-size results of the Berry phase for $N=12$ agree with the 'energy boundary' except for $0.5\leq J_2\leq0.6$. The finite-size extrapolation of the Berry phase is close to the 'dimerization boundary'.

\section{Solitons at the first order transition between Haldane and dimerized phases}
\label{sec:solitons}

We have studied numerically the soliton formation around the first order phase transition between the Haldane and dimerized phases. In Fig.\ref{fig:solitons} we show results for the lowest-lying $S^z_\mathrm{tot}=1$ states of a $N=61$ site chain for $J_2=0.3$ and different values of $J_3$. The most relevant quantities are: i) the local magnetization $\langle S^z_j\rangle$ that reveals edge states or solitons; ii) the spin-spin correlation between nearest neighbors $\langle S^z_j S^z_{j+1}\rangle$ that reflects the presence of dimerization; and iii) the expectation value of the three-site interaction $ \langle({\bf S}_{i-1}{\bf S}_{i})({\bf S}_{i}{\bf S}_{i+1})-\mathrm{h.c.} \rangle$, an indiactor of the Haldane phase - it is large and positive in the Haldane phase, it vanishes when the state is exactly dimerized and it is negative or vey small and positive everywhere else in dimerized phase.

Our main results can be summarized as follows: Deep inside the Haldane phase there are spin-1/2 edge states as seen from the local magnetization of Fig.\ref{fig:solitons}a. The small dimerization and the large expectation value of the three-body interaction all along the chain confirm that the entire chain is in the Haldane phase. 
Around the phase transition, two phases coexist: the dimerized state is favoured close to the edges, while the  central part of a chain remains in the Haldane phase. (Fig.\ref{fig:solitons}e-f). The two humps of the local magnetization curve (Fig.\ref{fig:solitons}d)  show that free spins have moved away from the boundaries and form a pair of spin-1/2 solitons that separates the Haldane and dimerized domains. 
Deep inside the dimerized phase, different dimerization domains are separated by spin-1 solitons (Fig.\ref{fig:solitons}i-k). The transition between two dimerization domains with different dimer orientations can also be deduced from the crossing of  the lines formed by red and blue points inthe  spin-spin correlation (see Fig.\ref{fig:solitons}j)


\begin{widetext}

\begin{figure}[t!]
\includegraphics[width=1\textwidth]{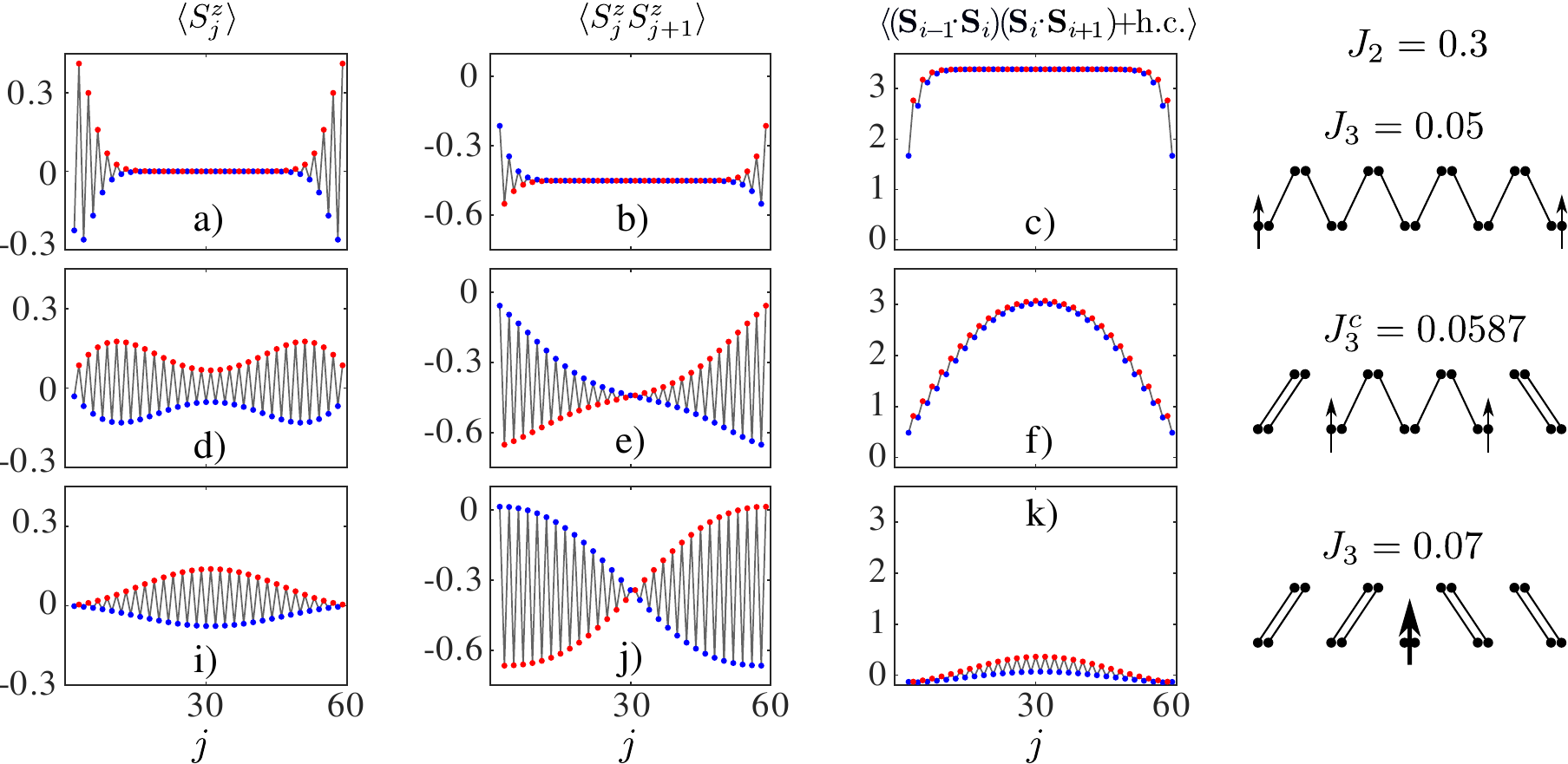}
\caption{(Color online) Spin solitons in chains with $N=61$ at $J_2=0.3$ and $S^z_\mathrm{tot}=1$ below (a-c), above (i-k) and on the critical line (d-f). Left panels: On-site magnetization. It reveals  a) spin-1/2 edge states, d) spin-1/2 solitons inside the chain, and i) spin-1 soliton. Middle panels: Spin-spin correlations. They provide evidence of a small dimerization all along the chain in the Haldane phase (b), of a large dimerization at the edges and of its fast decrease in the middle, when the two phases coexist (e), and of a large dimerization all along the chain except in the narrow window in the middle, where the spin-1 soliton is located (j). Right panel: Expectation value of the three-body term. It is large and positive all along the chain in he Haldane phase (c), it is small at the dimerized edges but remains large in a domain of Haldane phase in the middle of a chain (f), and it almost vanishes in the dimerized phase (k). The sketches on the right show the VBS picture of solitons in different phases. Thin and thick arrows indicates spin-1/2 and spin-1 solitons. For clarity, each even (odd) data point corresponds to a blue (red) symbol. }
\label{fig:solitons}
\end{figure}

\begin{figure}[h!]
\includegraphics[width=1\textwidth]{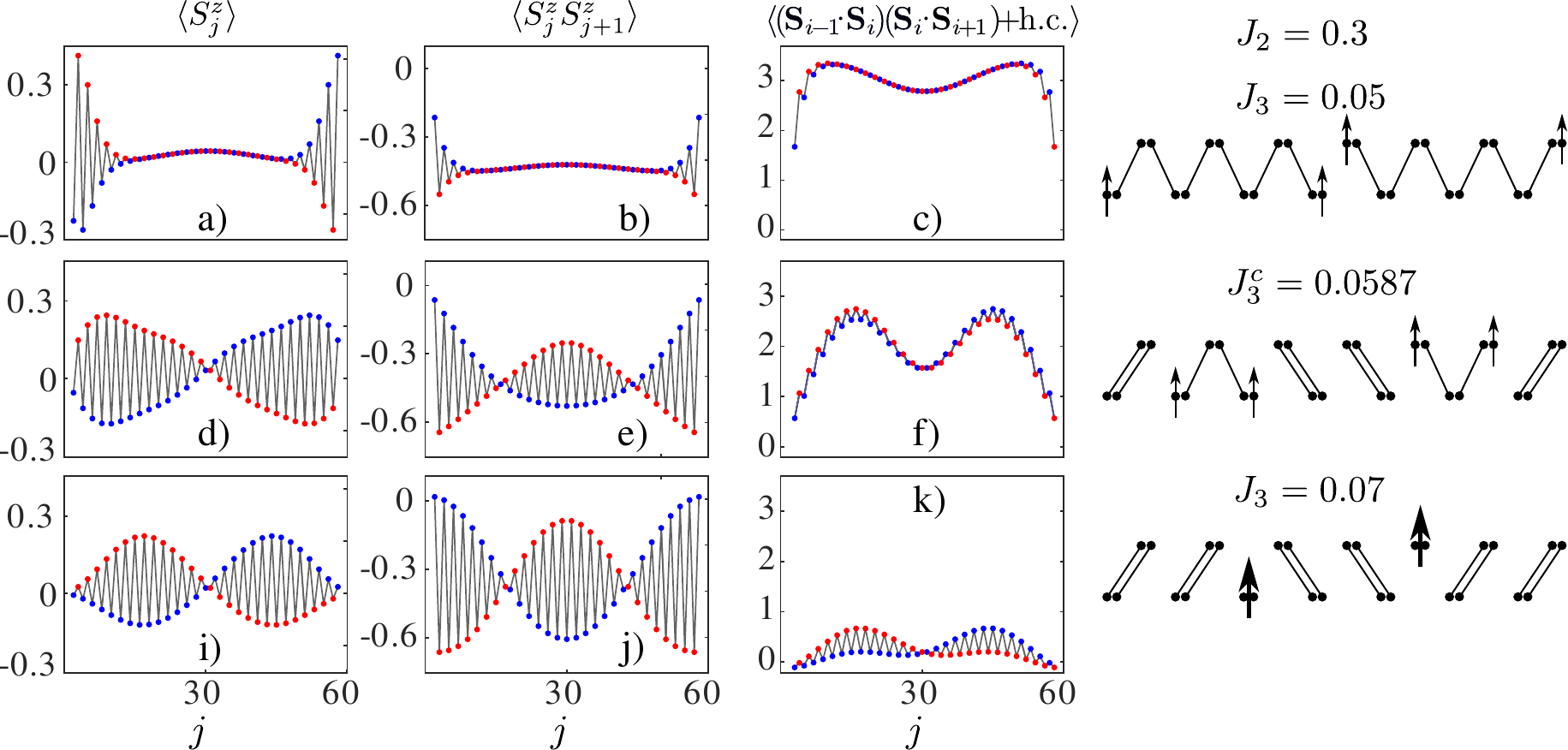}
\caption{(Color online) Same as Fig.\ref{fig:solitons} for $N=60$ at $J_2=0.3$ and $S^z_\mathrm{tot}=2$ (see main text for the details). }
\label{fig:solitons2}
\end{figure}

\end{widetext}

The soliton picture remains true for higher values of the total spin. Fig.\ref{fig:solitons2} provides an example of solitons in a chain with $N=60$ and $S^z_\mathrm{tot}=2$. 
As in the previous case, one can clearly distinguish spin-1/2 edge states in Fig.\ref{fig:solitons2}a). On top of it, a slight increase of the on-site magnetization occurs in the midlle of the chain, indicating the appearance of a spin-1 bond in the bulk. The Haldane phase is then perturbed. This is reflected in the suppressed three-body term measured in the middle of the chain (Fig.\ref{fig:solitons2}c).
The formation of the dimerized phase starts at the edges, but also in the bulk when approaching the phase transition. Different dimerization domains are separated by domains of Haldane phase, each carrying a total spin-1. Numerically, three dimerization domains are clearly seen with the spin-spin correlations in Fig.\ref{fig:solitons2}e), while the two maxima in the three-body term correspond to two Haldane domains.
Inside the dimerized phase, the Haldane domains are suppressed, and two spin-1 solitons separate the dimerization domains. The lines formed by red and blue points in the spin-spin correlation intersect twice in Fig.\ref{fig:solitons2}e,c), implying that the orientation of the dimers is different in neighboring domains.


\section{Ising Transition}
\label{sec:ising}

Previously, it was shown that the transition between the NNN-Haldane and dimerized phases is in the Ising universality class at a generic point on the transition line\cite{J1J2J3_letter}. In this section we numerically check that the universality class remains the same at the two edges of the transition line: at the triple point where three phases touch each other, and in the limit of large $J_2$ and $J_3$ couplings.

\subsection{Triple point}

There are two possible scenarios for the triple point of Ising critical line: it can be in either in the Ising or in the tricritical Ising universality class. According to conformal field theory, the first scenario is characterized by a scaling dimension $d=1/8$ and a central charge $c=1/2$, while the second one is characterized by $d=1/24$ and $c=7/10$. 

\begin{figure}[h!]
\includegraphics[width=0.45\textwidth]{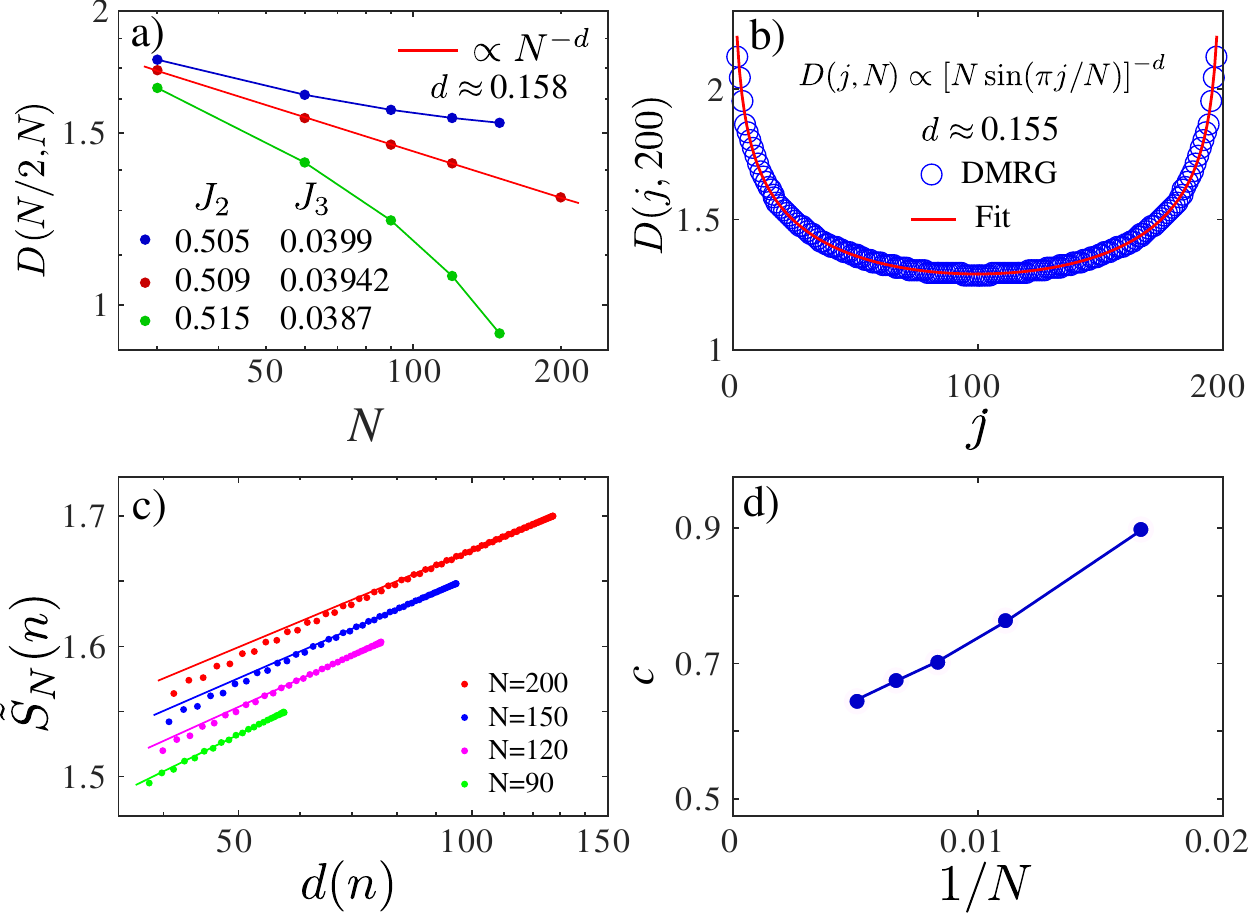}
\caption{(Color online) a) Log-log plot of the mid-chain dimerization as a function of the number of sites N for different parameters $J_2$ and $J_3$ along the line close to the first order phase transition and perpendicular to the Ising critical line. The linear curve corresponds to the critical point and the slope gives the critical exponent $d\approx 0.158$. b) Site dependence of $D(j,N)$ at the critical point fitted to $1/\left[N\sin(\pi j/N)\right]^d$. This determines the exponent $d\approx 0.155$. c) Scaling of the entanglement entropy of open chains after removing the Friedel oscillations with conformal distance $d(n)$. d) Central charge extracted from the entanglement entropy of open chains as a function of $1/N$}
\label{fig:ising_end_point}
\end{figure}

We have looked at the critical point along a line that is very close to the first order transition and perpendicular to the Ising critical line. According to conformal field theory, the local dimerization depends on the chain length $N$ and bond index $j$ as $D(j,N)=\left[N\sin(\pi j/N)\right]^{-d}$. The values of $J_2$ and $J_3$, for which the scaling of the mid-chain dimerization $D(N/2,N)$ is a separatrix is taken as the critical point (Fig.\ref{fig:ising_end_point}a). At the critical point the fit of $D(j,N)$ is also good (Fig.\ref{fig:ising_end_point}b). The resulting values of $d\approx 0.158$ and $0.155$ point rather towards Ising than towards tricritical Ising criticality.

The central charge was extracted at the critical point from the scaling of entanglement entropy with block size in open systems. Following Ref.\onlinecite{capponi}, we defined the reduced entanglement entropy $\tilde{S}_N(n)$ as the one with removed Friedel oscillations:
\begin{equation}
\tilde{S}_N(n)=S_N(n)-\zeta\langle {\bf S}_n{\bf S}_{n+1} \rangle,
\label{eq:calabrese_cardy_obc_corrected}
\end{equation}
where $\zeta$ is a numerical parameter.
Then, according to conformal field theory the reduced entanglement entropy scales with conformal distance $d(n)=\frac{2N}{\pi}\sin\left(\frac{\pi n}{N}\right)$ according to:
\begin{equation}
\tilde{S}_N(n)=\frac{c}{6}\ln d(n)+s_1+\log g
\label{eq:calabrese_cardy_obc}
\end{equation}
Although our numerical result point out to a central charge $c\approx 0.6$, that is in between the two expected values, the monotonous decrease and the fact that for $N=150,200$ the central charge is below $0.7$ suggest that the critical point is in the Ising universality class.

\subsection{$J_2-J_3$ model}

\begin{figure}[h!]
\includegraphics[width=0.45\textwidth]{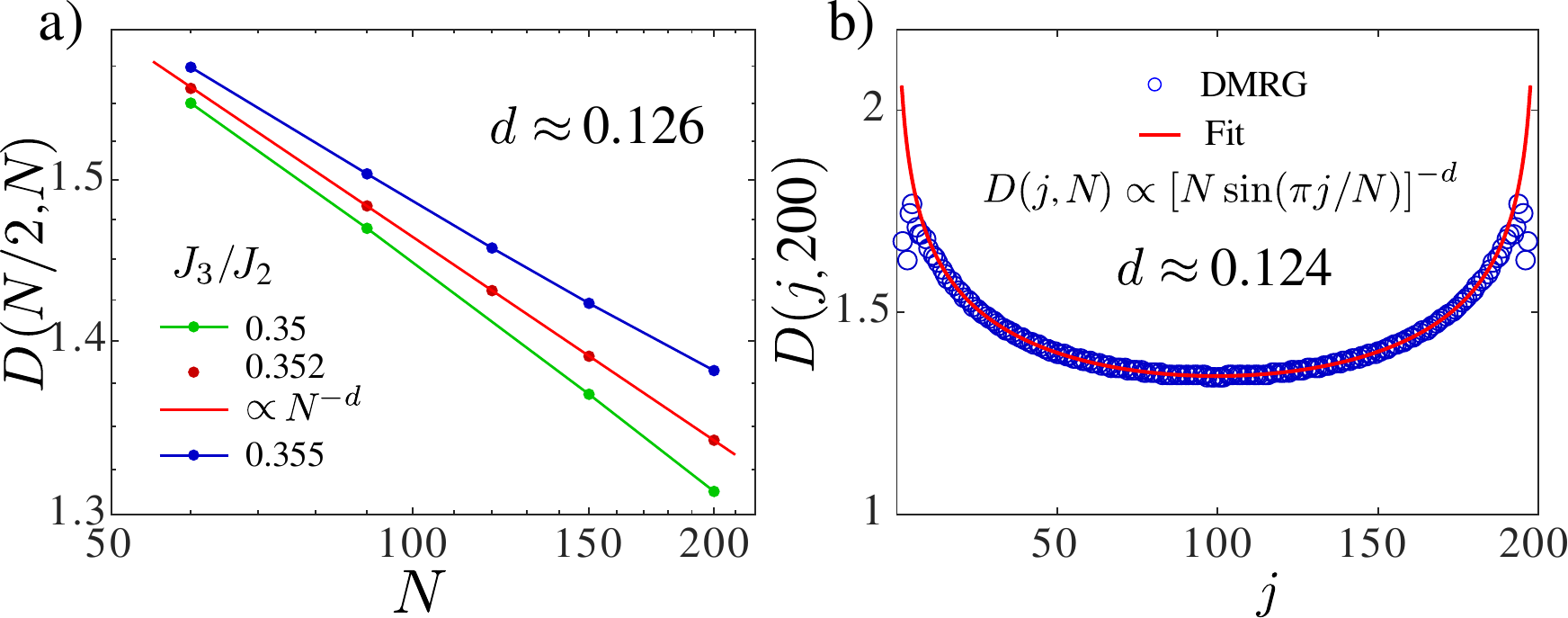}
\caption{(Color online) a) Log-log plot of the mid-chain dimerization as a function of the number of sites N for $J_1=0$, $J_2=1$ and different values of $J_3$. The linear curve corresponds to the Ising critical point and the slope gives a critical exponent $d=0.126$, in good agreement with $1/8$ for the Ising transition. b) Site dependence of $D(j,N)$ at the critical point fitted to $1/\left[N\sin(\pi j/N)\right]^d$. This leads to an exponent $d=0.124$, again close to the Ising prediction $1/8$}
\label{fig:J2J3}
\end{figure}

We have studied the limit of large $J_2$ and $J_3$ couplings by setting the nearest neighbor interaction to zero: $J_1=0$. As above, we locate the critical point by looking for the separatrix in the scaling of the mid-chain dimerization $D(N/2,N)$ with the chain length $N$. The slope gives a critical exponent $d\approx0.126$, in excellent agreement with the Ising one (see Fig.\ref{fig:J2J3}a). We also looked at the local dimerization $D(j,N)$ as a function of the bond position $j$. Although the dimerization remains large close to the boundary, one can clearly see that some edge effects appear in the absence of a $J_1$ coupling. A similar picture arises in the Ising chain in a transverse field if the up-up boundary field is weak with respect to the transverse field.  We have thus excluded a few edge points from the fit. The rest of the curve is again in excellent agreement with the Ising prediction $d=1/8$ (see Fig.\ref{fig:J2J3}b). 

\begin{figure}[h!]
\includegraphics[width=0.47\textwidth]{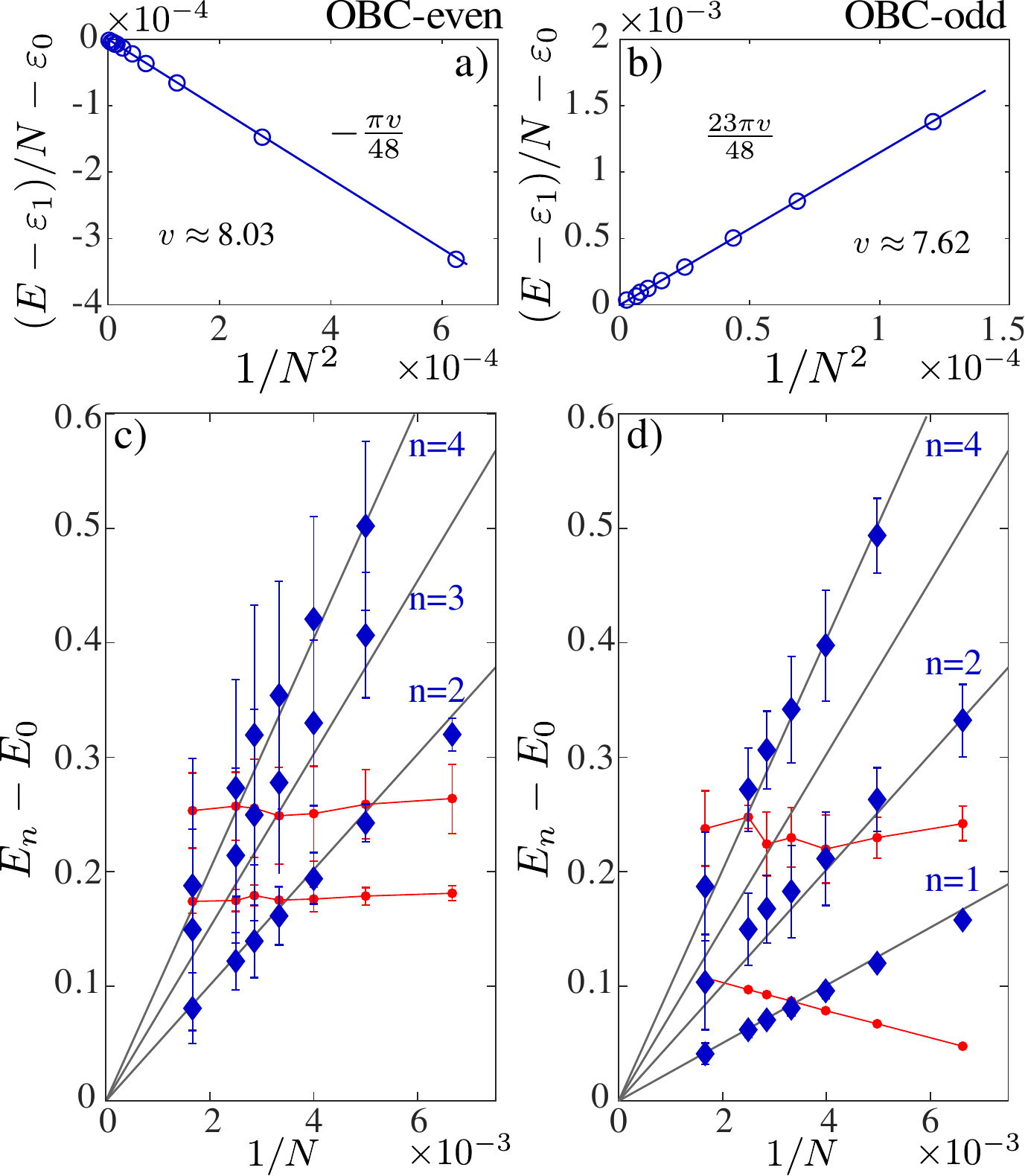}
\caption{(Color online) Ground state and excitation energy as $J_1=0$, $J_2=1$ and $J_3=0.352$, on the Ising line. a) and b) Linear scaling of the groundstate energy per site in an open chain with $1/N^2$ after subtracting the $\varepsilon_0$ and $\varepsilon_1$ terms for even (a) and odd (b) number of sites. c) and d) Energy gap in singlet (blue) and triplet (red) sectors for OBC as a function of $1/N$ for even and odd number of sites. Each magnetic excitation is twofold degenerate. Grey lines mark the Ising conformal towers of $I$ c) and of $\epsilon$ d) with the velocity $v=8.03$ deduced from the finite-size scaling of the ground state energy for even $N$. }
\label{fig:j2j3_ct}
\end{figure}

As predicted by boundary conformal field theory, for the Ising critical point the ground-state energy of an open system with an even number of sites scales as $E=\varepsilon_0N+\varepsilon_1-\pi v/(48N)$, where $\varepsilon_0$ is a groundstate energy per site, $\varepsilon_1$ - is a non-universal constant, and $v$ is the velocity. For odd $N$ the scaling is of the form $E=\varepsilon_0N+\varepsilon_1+23\pi v/(48N)$. We present the fit of the numerical data in the Fig.\ref{fig:j2j3_ct}a-b). The extracted values of the velocities $v_\mathrm{even}\approx8.03$ and $v_\mathrm{odd}\approx7.62$ are in reasonable agreement with each other.

 As discussed in [\onlinecite{J1J2J3_letter}], if the formation of a dimer is favoured on the edge bonds, the chain with even (odd) number of sites $N$ is identified with $\uparrow,\uparrow$ ($\uparrow,\downarrow$) boundary conditions in the Ising model. Then the conformal tower for even $N$ corresponds to the Ising conformal tower of $I$, while for odd $N$ it corresponds to the Ising conformal tower of $\epsilon$. We have used the velocity $v_\mathrm{even}$ deduced from the finite-size scaling of the ground state energy for even $N$ in order to plot the Ising towers of $I$ and of $\epsilon$ in Fig.\ref{fig:j2j3_ct}c-d) as references.
 
We have calculated the excitation energy for even and odd numbers of sites in the singlet $S^z_\mathrm{tot}=0$ and triplet $S^z_\mathrm{tot}=1$ sectors. The absence of the $J_1$ term releases low-lying magnetic excitations that are shown with red lines in Fig.\ref{fig:j2j3_ct}c-d). Each red line is twofold degenerate, corresponding to the excitation close to the left and to the right edges. The first singlet excitation appears below the triplet one only for $N>300$. By looking at the excitation energy as a function of the number of DMRG iterations, or, more specifically, as a function of the position of the state tensor updated at each iteration, we were able to distinguish bulk excitations from the excitations at the edges, even when they were above the first triplet excitation. Note that by edge excitation we understand a local magnetic excitation of a bond that is located close to the chain boundary. 

Since the calculations had to be done for very large systems (in Fig.\ref{fig:j2j3_ct}c-d),  we present the results for $N$ in the range $150$ to $601$.), the convergence of the algorithm is quite slow, implying significant error bars. 
In systems with non-zero $J_1$ coupling, we saw that for an odd number of sites the fourth excitation was more stable in the DMRG sense than the third one. This explains the 'missing' third excitation on panel d): we were not able to converge enough excited states for systems that are so large.

To summarize, we have have provided numerical evidence that the phase transition between the NNN-Haldane and dimerized phases is always in the Ising universality class, including at the triple point where the Haldane, NNN-Haldane and dimerized phases touch, and in the limiting case of the $J_2-J_3$ model.


\section{Short Range Order}
\label{sec:short_range_order}

\subsection{Disorder and Lifshitz lines}

\begin{figure}[h!]
\includegraphics[width=0.47\textwidth]{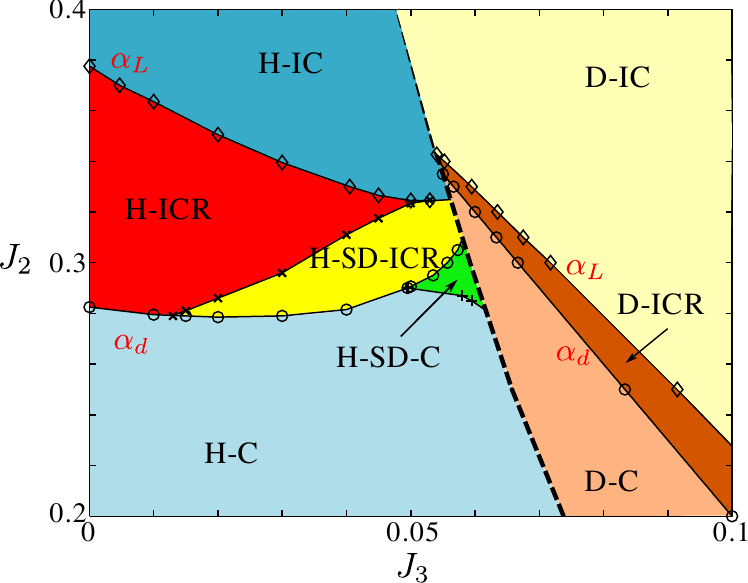}
\caption{(Color online) Enlarged part of phase diagram on Fig.\ref{fig:PD_short} indicating short-range order within the thermodynamic phases. Lifshitz line $\alpha_L$ is marked with diamonds and disorder line $\alpha_d$ is marked with open circles. Dashed line is a line of the first order phase transition}
\label{fig:ShortRangeDiag2}
\end{figure}

As mentioned in Section \ref{sec:phase_diagram}, several types of short-range order are present in the Haldane, NNN-Haldane and dimerized phases (see Fig.\ref{fig:ShortRangeDiag2}). A detailed description of each phase has already be given in Section \ref{sec:phase_diagram}. In this section, we describe the numerical results that led to this phase diagram in more detail. 

The most important result is that, by tuning either the next-nearest-neighbor or the three-body interaction, short-range incommensurate order can be induced beyond the so-called disorder and Lifshitz lines. 
Disorder points were first discussed by Stephenson in models of classical statistical mechanics\cite{stephenson1,stephenson2,stephenson3}. On one side of a disorder point, the correlation function decays in a commensurate way, while on the other side it decays in an incommensurate way. The disorder point is said to be of the first kind if the wave number in the incommensurate phase depends on the temperature, and of the second kind if it does 
not\cite{stephenson2}. In the present case, we have only found disorder points of the first kind.

By contrast, at a Lifshitz transition, the spin-spin correlation function becomes incommensurate in momentum space, each peak being replaced  by two symmetric peaks in the structure factor $SF(q)$ defined in Eq.\ref{eq:sf}.

By keeping track of real space and momentum space correlations, we found that disorder and Lifshitz lines cross the transition line at $J_2\simeq0.335$ and $J_2\simeq0.342$. 

\subsection{Dimerized phase}

\begin{figure}[h!]
\includegraphics[width=0.45\textwidth]{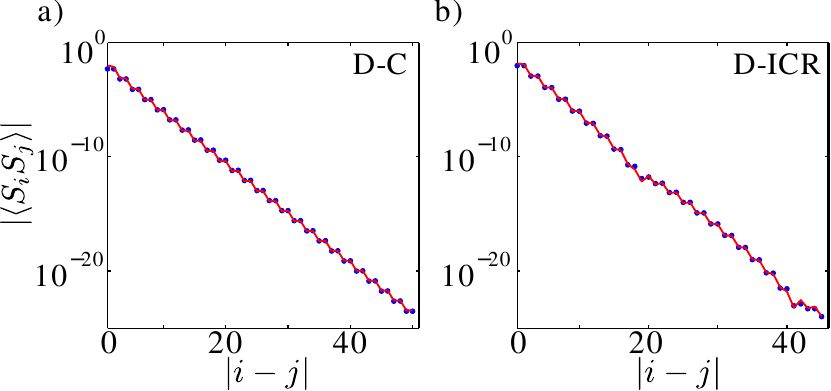}
\caption{(Color online) Spin-spin correlation function $\langle S_iS_j\rangle$ for $J_2=0.25$ with a) $J_3=0.083$ and b) $J_3=0.084$; The red lines are fit to the data with the dimerized OZ form given by Eq.\ref{eq:OZD} with a) $q=\pi$ and b) $q>\pi$}
\label{fig:short_range_cor_dim}
\end{figure}

By fitting the numerical results of the spin-spin correlations with the dimerized OZ form given by Eq.\ref{eq:OZD}, we have extracted the wave number $q$ and the short-range dimerization parameter $\delta$. Examples of fits for $J_2=0.25$ are shown in Fig.\ref{fig:short_range_cor_dim}.

\begin{figure}[h!]
\includegraphics[width=0.47\textwidth]{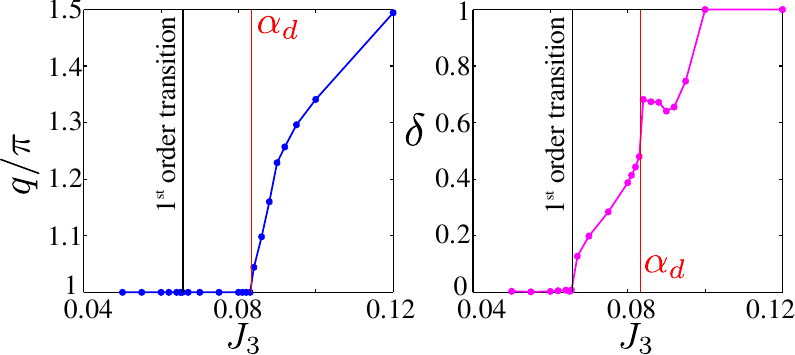}
\caption{(Color online) Wave number $q$ and dimerization $\delta$ deduced from a fit of the spin-spin correlation function with Eq.\ref{eq:OZD} for $J_2=0.25$. The position of the first order phase transition and the line $\alpha_d$ where the ground state is fully dimerized are marked with black and red lines respectively.}
\label{fig:ShortRange_Dim}
\end{figure}

We have found that, with very high accuracy, the disorder line coincides with the line where the fully dimerized wave-function is the exact 
ground state of the model (see Fig.\ref{fig:ShortRange_Dim}).

\begin{figure}[h!]
\includegraphics[width=0.4\textwidth]{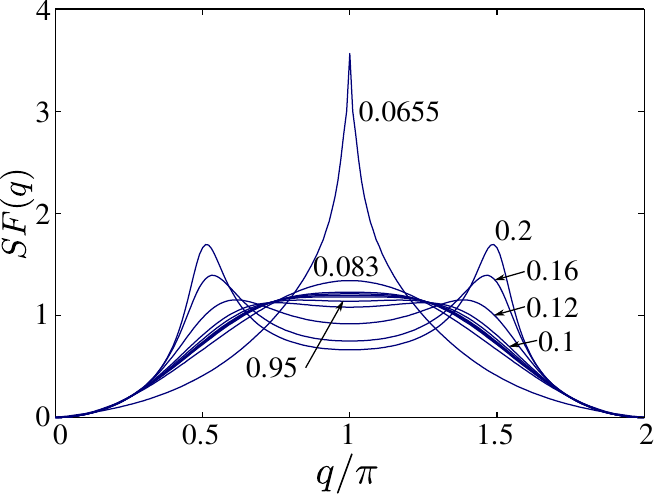}
\caption{(Color online) Structure factor $SF(q)$ for $J_2=0.25$ and various values of $J_3$. The Lifshitz point is at $J_3=0.0915\pm0.0005$.}
\label{fig:StructFact}
\end{figure}

In order to determine the Lifshitz line, we have looked for the appearance of a two-peak structure in $SF(q)$ given by Eq.\ref{eq:sf}, where we have restricted the sum to the interval $20<i,j\leq N-20$ in order to eliminate edge effects. Some examples of structure factor calculated for fixed $J_2=0.25$ are presented in Fig.\ref{fig:StructFact}.

The conclusion that emerges from these results is that, to go from the commensurate to the incommensurate part of the dimerized phase, one has to cross first a disorder line, and then a Lifshitz line.
These results are very similar to those obtained for the spin-1/2 chain with next-nearest-neighbor interaction, the
fully-dimerized line of our model being the equivalent of the Majumdar-Ghosh point\cite{MajumdarGhosh}. At that point, the correlation length vanishes,  and it coincides with the disorder point $J_2^d=1/2$, while the Lifshitz point of the spin-1/2 chain is located at $J_2^L=0.52036(6)$\cite{bursill}, well above the disorder point.

\subsection{Haldane phase}

Depending on the type of correlation in the Haldane phase we have fitted the numerical data with either non-dimerized OZ or dimerized OZ forms given by Eq.\ref{eq:OZ} and Eq.\ref{eq:OZD}. Below we provide several examples of spin-spin correlations and some fits.

\begin{figure}[h!]
\includegraphics[width=0.47\textwidth]{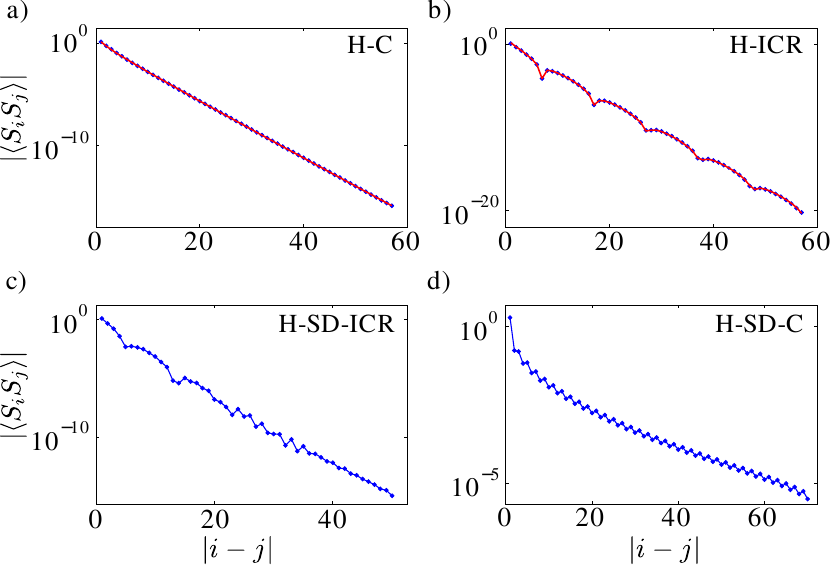}
\caption{(Color online) Spin-spin correlation function $\langle S_iS_j\rangle$ for a) $J_2=0.27$, $J_3=0$; b) $J_2=0.3$, $J_3=0$; c) $J_2=0.3$, $J_3=0.04$ and d) $J_2=0.3$, $J_3=0.059$. The red line on a) and b) is a fit to the data with the OZ form.}
\label{fig:CC_endpoint}
\end{figure}

\begin{figure}[h!]
\includegraphics[width=0.47\textwidth]{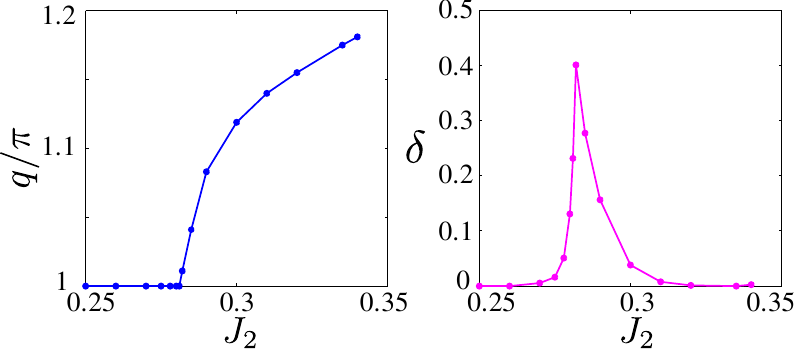}
\caption{(Color online) Wave number $q$ and dimerization $\delta$ deduced from a fit of the spin-spin correlation function with Eq.\ref{eq:OZD} for $J_3=0.03$. }
\label{fig:ShortRange_Hald}
\end{figure}

The wave number $q$ and the short-range dimerization parameter $\delta$ extracted from the fit for fixed $J_3=0.03$ are summarized in Fig.\ref{fig:ShortRange_Hald}. Note that there is a finite region where the dimerization is essentially different from zero.

Crossing the transition line at $J_2\simeq0.335$, the disorder line is separated from the transition line in Haldane phase by a thin tail of commensurate phase with short-order dimerization (H-SD-C).

The Lifshitz line in the Haldane phase is obtained in the same way as in the dimerized phase. Close to the crossing point $J_2\simeq0.342$, the Lifshitz line is very close to the boundary of the H-SD-ICR phase, making the H-ICR phase vanishingly small in this region.


\section{Conclusion} 
\label{sec:conclusion}

Combining field theory arguments with DMRG (and occasionally exact diagonalizations), we have shown that the dimerization transitions 
of the spin-1 Heisenberg model with next-nearest neighbor and three-site interaction can be precisely located and fully characterized. 
In particular, the transition between the Haldane phase and the dimerized phase is in the $SU(2)_2$ WZW universality class for small $J_2$, and it becomes
first order at an end point also in the $SU(2)_2$ WZW universality class. Along the first order line between these phases, the solitons between
Haldane and dimerized phases carry a spin-1/2, in qualitative agreement with the fact that along the $SU(2)_2$ line, there are low-lying magnetic
excitations. By contrast, the transition between the next-nearest neighbor Haldane phase and the dimerized phase is in the Ising universality
class. Along this transition line, the spin-gap remains open, and the low-lying excitations are all in the singlet sector. 

To fully characterize the transitions, DMRG with open boundary conditions turned out to be extremely useful. This is due to the fact that
the conformal tower of a critical model with open boundary conditions is often just the tower of a single primary field. By contrast, the conformal tower
of a critical model with periodic boundary conditions is in general the superposition of different towers. We think that a systematic use of these ideas
might turn out to be useful in other one-dimensional quantum systems.

In addition, we have shown that short-range correlations can be commensurate or incommensurate, with several disorder and Lifshitz lines, leading to a remarkably rich phase diagram. Interestingly, several of these phases occur for relatively small, hence physically realistic values of the couplings $J_2$ and $J_3$. So it is our hope that the present investigation will encourage experimentalists to try and check some aspects of this phase diagram.

\section{Acknowledgments} 

We are indebted to Andreas L\"auchli, and Andrey Nevidomskyy for insightful discussions and advises. The first evidence of a partially
first order transition between the Haldane phase and the dimerized phase has been obtained by Cl\'ement Bazin during his Master thesis. 
This work has been supported by the Swiss National Science Foundation and by NSERC of Canada, Discovery Grant 04033-2016 (IA) and the Canadian Institute for Advanced Research (IA).

\bibliographystyle{prsty}
\bibliography{bibliography}

\end{document}